\documentclass[12pt,graphicx,subfigure,axodraw]{article}
\usepackage{amsfonts}
\usepackage{amssymb}
\usepackage[usenames,dvipsnames]{color}
\usepackage{graphicx}
\usepackage{amsmath}
\usepackage{amsfonts}
\usepackage{amssymb}
\usepackage{cleveref}
\usepackage{float}
\restylefloat{table}
\usepackage{chngpage}
\def\rmuu{\gamma^{\mu}}
\def\rmud{\gamma_{\mu}}
\def\PL{{1-\gamma_5\over 2}}
\def\PR{{1+\gamma_5\over 2}}
\def\sinW2{\sin^2\theta_W}
\def\AEM{\alpha_{EM}}
\def\mul{M_{\tilde{u} L}^2}
\def\mur{M_{\tilde{u} R}^2}
\def\mdl{M_{\tilde{d} L}^2}
\def\mdr{M_{\tilde{d} R}^2}
\def\mz2{M_{z}^2}
\def\c2b{\cos 2\beta}
\def\au{A_u}
\def\ad{A_d}
\def\cob{\cot \beta}
\def\v#1{v_#1}
\def\tb{\tan\beta}
\def\epem{$e^+e^-$}
\def\KK{$K^0$-$\overline{K^0}$}
\def\wi{\omega_i}
\def\xj{\chi_j}
\def\Wmu{W_\mu}
\def\Wnu{W_\nu}
\def\m#1{{\tilde m}_#1}
\def\mH{m_H}
\def\mw#1{{\tilde m}_{\omega #1}}
\def\mx#1{{\tilde m}_{\chi^{0}_#1}}
\def\mc#1{{\tilde m}_{\chi^{+}_#1}}
\def\mwi{{\tilde m}_{\omega i}}
\def\mxi{{\tilde m}_{\chi^{0}_i}}
\def\mci{{\tilde m}_{\chi^{+}_i}}

\def\ch{{\tilde\chi^{+}_1}}
\def\c2{{\tilde\chi^{+}_2}}

\def\tt{{\tilde\theta}}

\def\tp{{\tilde\phi}}

\def\mz{M_z}
\def\sw{\sin\theta_W}
\def\cw{\cos\theta_W}
\def\cb{\cos\beta}
\def\sb{\sin\beta}
\def\rwi{r_{\omega i}}
\def\rxj{r_{\chi j}}
\def\rfp{r_f'}
\def\Kik{K_{ik}}
\def\Fq2{F_{2}(q^2)}
\def\f{\({\cal F}\)}
\def\d1{{\f(\tilde c;\tilde s;\tilde W)+ \f(\tilde c;\tilde \mu;\tilde W)}}
\def\tw{\tan\theta_W}
\def\sec2w{sec^2\theta_W}
\begin{document}
\baselineskip 18pt
\def\today{\ifcase\month\or
 January\or February\or March\or April\or May\or June\or
 July\or August\or September\or October\or November\or December\fi
 \space\number\day, \number\year}
\def\thebibliography#1{\section*{References\markboth
 {References}{References}}\list
 {[\arabic{enumi}]}{\settowidth\labelwidth{[#1]}
 \leftmargin\labelwidth
 \advance\leftmargin\labelsep
 \usecounter{enumi}}
 \def\newblock{\hskip .11em plus .33em minus .07em}
 \sloppy
 \sfcode`\.=1000\relax}
\let\endthebibliography=\endlist
\def\lsim{\ ^<\llap{$_\sim$}\ }
\def\gsim{\ ^>\llap{$_\sim$}\ }
\def\r2{\sqrt 2}
\def\beq{\begin{equation}}
\def\eeq{\end{equation}}
\def\beqn{\begin{eqnarray}}
\def\eeqn{\end{eqnarray}}
\def\rmuu{\gamma^{\mu}}
\def\rmud{\gamma_{\mu}}
\def\PL{{1-\gamma_5\over 2}}
\def\PR{{1+\gamma_5\over 2}}
\def\sinW2{\sin^2\theta_W}
\def\AEM{\alpha_{EM}}
\def\mul{M_{\tilde{u} L}^2}
\def\mur{M_{\tilde{u} R}^2}
\def\mdl{M_{\tilde{d} L}^2}
\def\mdr{M_{\tilde{d} R}^2}
\def\mz2{M_{z}^2}
\def\c2b{\cos 2\beta}
\def\au{A_u}
\def\ad{A_d}
\def\cob{\cot \beta}
\def\v#1{v_#1}
\def\tb{\tan\beta}
\def\epem{$e^+e^-$}
\def\KK{$K^0$-$\bar{K^0}$}
\def\wi{\omega_i}
\def\xj{\chi_j}
\def\Wmu{W_\mu}
\def\Wnu{W_\nu}
\def\m#1{{\tilde m}_#1}
\def\mH{m_H}
\def\mw#1{{\tilde m}_{\omega #1}}
\def\mx#1{{\tilde m}_{\chi^{0}_#1}}
\def\mc#1{{\tilde m}_{\chi^{+}_#1}}
\def\mwi{{\tilde m}_{\omega i}}
\def\mxi{{\tilde m}_{\chi^{0}_i}}
\def\mci{{\tilde m}_{\chi^{+}_i}}
\def\mz{M_z}
\def\sw{\sin\theta_W}
\def\cw{\cos\theta_W}
\def\cb{\cos\beta}
\def\sb{\sin\beta}
\def\rwi{r_{\omega i}}
\def\rxj{r_{\chi j}}
\def\rfp{r_f'}
\def\Kik{K_{ik}}
\def\Fq2{F_{2}(q^2)}
\def\f{\({\cal F}\)}
\def\d1{{\f(\tilde c;\tilde s;\tilde W)+ \f(\tilde c;\tilde \mu;\tilde W)}}
\def\tw{\tan\theta_W}
\def\sec2w{sec^2\theta_W}
\def\ch{{\tilde\chi^{+}_1}}
\def\c2{{\tilde\chi^{+}_2}}

\def\tt{{\tilde\theta}}

\def\tp{{\tilde\phi}}

\def\mz{M_z}
\def\sw{\sin\theta_W}
\def\cw{\cos\theta_W}
\def\cb{\cos\beta}
\def\sb{\sin\beta}
\def\rwi{r_{\omega i}}
\def\rxj{r_{\chi j}}
\def\rfp{r_f'}
\def\Kik{K_{ik}}
\def\Fq2{F_{2}(q^2)}
\def\f{\({\cal F}\)}
\def\d1{{\f(\tilde c;\tilde s;\tilde W)+ \f(\tilde c;\tilde \mu;\tilde W)}}

\def\b{${\cal{B}}(\mu\to {e} \gamma)$~}

\def\tw{\tan\theta_W}
\def\sec2w{sec^2\theta_W}
\newcommand{\pn}[1]{{\color{red}{#1}}}

\begin{titlepage}

\begin{center}
{\large {\bf
 Chromoelectric Dipole Moments of Quarks in MSSM Extensions
}}\\
\vspace{2cm}
\renewcommand{\thefootnote}
{\fnsymbol{footnote}}
Amin Aboubrahim$^{b}$,  Tarek Ibrahim$^{a}$\footnote{Email: tibrahim@zewailcity.edu.eg},
   Pran Nath$^{c}$\footnote{Email: nath@neu.edu} and Anas Zorik$^{d}$$\footnote{Email: anas.zorik@alexU.edu.eg}$
\vskip 0.5 true cm
\end{center}

\date{Feb 14, 2015}

\noindent
{$^{a}$University of Science and Technology, Zewail City of Science and Technology,}\\
{ 6th of October City, Giza 12588, Egypt\footnote{Permanent address:  Department of  Physics, Faculty of Science,
University of Alexandria, Alexandria 21511, Egypt}\\
}
{$^{b}$Department of Physics, Faculty of Science, Beirut Arab University,
Beirut 11-5020, Lebanon\footnote{Email: abouibrahim.a@husky.neu.edu}} \\
{$^{c}$Department of Physics, Northeastern University,
Boston, MA 02115-5000, USA} \\
{$^{d}$Department of Physics, Faculty of Science, Alexandria University, Alexandria 21511, Egypt}
\vskip 1.0 true cm

\centerline{\bf Abstract}
An analysis is given of the chromoelectric dipole moment of quarks and of the neutron in an MSSM
extension where the matter sector contains an extra vectorlike generation of quarks and mirror quarks.
The analysis includes contributions to the CEDM from the exchange of the $W$ and the $Z$ bosons,
from the exchange of charginos and neutralinos and the gluino. Their contribution to the EDM of quarks
is investigated.
{The interference between the MSSM sector and the new sector with vectorlike quarks is investigated.
It is shown that inclusion of the vectorlike quarks can modify the quark EDMs in a significant way.
Further, this interference also provides a  probe of the vectorlike quark sector.}
These results are of interest as in the future measurements on the neutron EDM could see an improvement
up to two orders of magnitude over the current experimental limits {and provide an
instrument for a further probe of new physics
beyond the standard model.}\\


\medskip

\end{titlepage}

\section{Introduction \label{sec1}}
 New sources of CP violation beyond those that exist in the Standard Model are needed to
 explain baryogenesis  and are also worthy of study in their own right as possible probes
 of beyond the standard model physics (for reviews
 see e.g.,~\cite{Golub:1994cg,sm,Ibrahim:2007fb,Hewett:2012ns}).
 Such sources can also induce electric dipole moment
 in elementary particles which can be significantly larger than those expected in the standard
 model~\cite{Golub:1994cg,sm}.
 In this work we are specifically interested in the electric dipole moment (EDM) of the quarks arising from the
 chromoelectric dipole operator. Thus the electroweak sector of the standard model produces
 an EDM which is  $10^{-30}$ ecm~\cite{Hoogeveen:1990cb,Soni:1992tn,note} and it lies beyond
 the possibility of its observation in the foreseeable future. As mentioned in particle physics models
 beyond the standard model it is  possible to generate much larger values for the EDM.
 In this work we focus on one such model - an extension of the minimal supersymmetric standard model
 (MSSM) with a vectorlike multiplet~\cite{vectorlike}. Such an extension is anomaly free and thus the nice quantum properties of MSSM are
 maintained. Further, vectorlike multiplets arise in a variety of settings such as in grand unified
 models and in string and D brane models~\cite{vectorlike,Liu:2009cc,Martin:2009bg}.
 Vectorlike generations have been considered by several authors since their discovery
  would constitute new physics (see, e.g., \cite{Babu:2008ge,Liu:2009cc,Martin:2009bg,Ibrahim:2011im,Ibrahim:2010hv,Ibrahim:2010va,Ibrahim:2008gg,Ibrahim:2009uv,Ibrahim:2012ds,Ibrahim:2014tba,Ibrahim:2015hva,Aboubrahim:2014hya,Aboubrahim:2013yfa,Aboubrahim:2013gfa}).
 Such models have new sources of CP violation
 and thus can generate substantial size dipole operators. For that reason they are interesting models to consider in the context of lepton and quark EDMs.
  In a recent work we analyzed the electric dipole operator
 in such a setting~\cite{Aboubrahim:2015nza} and in this work we analyze the chromoelectric dipole operator in the extended MSSM model and its contribution to the electric dipole moments. \\

  Before discussing the EDM in the new class of models, it is relevant to recall the situation regarding the lepton and
  quark EDMs in MSSM. Here it is well known that MSSM has a SUSY CP problem, i.e., that the EDM predicted
  with SUSY phases $O(1)$ are typically in excess of the experimental upper limits. A number of remedies have
  been offered in the past to remedy this problem. These include a fine tuning of the phases to be small~\cite{earlywork},
  suppression of the EDM by large sparticle masses~\cite{Nath:1991dn},  suppression of the EDM where various
  contributions conspire to cancel, i.e., the cancellation mechanism~\cite{cancellation1,cancellation2} as well as other
  possible remedies (see, e.g., ~\cite{Babu:1999xf}). It has also been suggested
  that the EDM be used as a probe of new physics beyond the standard model ~\cite{McKeen:2013dma,Moroi:2013sfa,Altmannshofer:2013lfa,Ibrahim:2014tba,Dhuria:2014fba}. Specifically the experimental limits on  the EDMs can be used as vehicles to probe
  a new physics regime not accessible otherwise to current and future detectors.

The outline of the rest of the paper is as follows: In section 2
 we give a brief description of the model
and describe the nature of mixing between the vector generation and the standard three generations of
quarks. In  section 3.1 we discuss the loop contributions to the chromoelectric dipole moment of the up quark
and the down quark that arise from the exchange of the $W$  boson in the loop. In section 3.2
we give an analysis similar to that of  section 3.1 for the exchange of the $Z$ boson. In section 3.3
we compute the contribution from the exchange of charginos in the loop and in section 3.4 a similar analysis
for the exchange of  neutralinos in the loop is given. Finally in  section 3.5 we give the analysis for the
exchange of the gluino in the loop.  In section 4 we discuss the method
for the computation of the neutron dipole moment using the quark dipole moments. In  section 5 we give
a detailed numerical analysis of the contributions to the quark CEDM and to the neutron CEDM for
a variety of parameter points in the extended MSSM model. Here we also discuss the use of the neutron
EDM as a probe of high mass scales. Conclusions are given in section 6. Further details of the
calculational aspects of the analysis are given in sections (7-9).

\section{The Model\label{sec2}}
Here we briefly describe the model and further details are given in the appendix.  The model we consider is an
extension of MSSM with an additional vectorlike multiplet. Like MSSM the vectorlike extension is free of anomalies
and as discussed in section 1 vectorlike multiplets appear in a variety of settings which include grand unified models,
string and D brane models. Here we focus on the quark sector where the vectorlike multiplet consists of a
 fourth generation of quarks and their mirror quarks.
 Thus the quark sector of the extended MSSM model is given by Eq. (1) and Eq. (2) where,

\begin{align}
q_{iL}\equiv
 \left(\begin{matrix} t_{i L}\cr
 ~{b}_{iL}  \end{matrix} \right)  \sim \left(3,2,\frac{1}{6}\right) \ ;  ~~ ~t^c_{iL}\sim \left(3^*,1,-\frac{2}{3}\right)\ ;
 ~~~ b^c_{i L}\sim \left(3^*,1,\frac{1}{3}\right)\ ;
  ~~~i=1,2,3,4.
\label{2}
\end{align}

\begin{align}
Q^c\equiv
 \left(\begin{matrix} B_{ L}^c \cr
 T_L^c\end{matrix}\right)  \sim \left(3^*,2,-\frac{1}{6}\right)\ ;
~~  T_L \sim  \left(3,1,\frac{2}{3}\right)\ ;  ~~   B_L \sim \left(3^*,1,-\frac{1}{3}\right).
\label{3}
\end{align}
The numbers in the braces show  the properties  under $SU(3)_C\times SU(2)_L\times U(1)_Y$
where the first two entries label the representations for $SU(3)_C$ and $SU(2)_L$ and the last one
gives the value of the hypercharge normalized so  that $Q=T_3+Y$.
We allow the mixing of the vectorlike generation with the first three generations. Specifically we will focus
on the mixings of the mirrors in the vectorlike generation with the first three generations. Details of these
mixings are given in Eq. (43). Here we display some relevant features.  In the up quark sector we
choose a basis as follows

\begin{gather}
\bar\xi_R^T= \left(\begin{matrix}\bar t_{ R} & \bar T_R & \bar c_{ R}
&\bar u_{R} &\bar t_{4R} \end{matrix}\right),~~
\xi_L^T= \left(\begin{matrix} t_{ L} &  T_L &  c_{ L}
& u_{ L} &\bar t_{4L}\end{matrix}\right)\,.
\label{basis-xi}
\end{gather}
and we write the mass term  so that

\beq
-{\cal L}^u_m= \bar\xi_R^T (M_u) \xi_L
+\text{h.c.},
\eeq
The interaction of Eq. (43) lead to the up-quark  mass matrix $M_u$
which is given by

\beqn
M_u=
 \left(\begin{matrix} y'_1 v_2/\sqrt{2} & h_5 & 0 & 0&0 \cr
 -h_3 & y_2 v_1/\sqrt{2} & -h_3' & -h_3''&-h_6 \cr
0&h_5'&y_3' v_2/\sqrt{2} & 0 &0\cr
0 & h_5'' & 0 & y_4' v_2/\sqrt{2}&0 \cr
0&h_8&0&0&y_5'v_2/\sqrt{2}\end{matrix}\right)\,.
\label{7aaa}
\eeqn
This mass matrix is not hermitian and a  bi-unitary transformation is needed  to diagonalize it.
Thus one has
\beq
D^{u \dagger}_R (M_u) D^u_L=\text{diag}(m_{u_1},m_{u_2},m_{u_3}, m_{u_4},  m_{u_5} ).
\label{7a}
\eeq
Under the bi-unitary transformations the basis vectors transform so that
\beqn
 \left(\begin{matrix} t_{R}\cr
 T_{ R} \cr
c_{R} \cr
u_{R} \cr
t_{4R}
\end{matrix}\right)=D^{u}_R \left(\begin{matrix} u_{1_R}\cr
 u_{2_R}  \cr
u_{3_R} \cr
u_{4_R}\cr
u_{5_R}
\end{matrix}\right), \  \
\left(\begin{matrix} t_{L}\cr
 T_{ L} \cr
c_{L} \cr
u_{L}\cr
t_{4L}
\end{matrix} \right)=D^{u}_L \left(\begin{matrix} u_{1_L}\cr
 u_{2_L} \cr
u_{3_L} \cr
u_{4_L}\cr
u_{5_L}
\end{matrix}\right) \ .
\label{8}
\eeqn

A similar analysis can be carried out for the down quarks. Here we choose the basis set as
\begin{gather}
\bar\eta_R^T= \left(\begin{matrix}\bar{b}_R & \bar B_R & \bar{s}_R
&\bar{d}_R
&\bar{b}_{4R}
\end{matrix}\right),
~~\eta_L^T= \left(\begin{matrix} {b_ L} &  B_L &  {s_ L}
& {d_ L}
&{b_{4L}}
 \end{matrix}\right)\,.
\label{basis-eta}
\end{gather}

In this basis  the down quark mass terms are given by
\beq
-{\cal L}^d_m=
\bar\eta_R^T(M_{d}) \eta_L
+\text{h.c.},
\eeq
where using the interactions of Eq. (43), $M_d$ has the following form
\beqn
M_d=\left(\begin{matrix} y_1 v_1/\sqrt{2} & h_4 & 0 & 0  & 0\cr
 h_3 & y'_2 v_2/\sqrt{2} & h_3' & h_3'' &h_6\cr
0&h_4'&y_3 v_1/\sqrt{2} & 0&0 \cr
0 & h_4'' & 0 & y_4 v_1/\sqrt{2}&0\cr
0& h_7 & 0 &0 &y_5 v_1/\sqrt{2}
\end{matrix} \right)\ .
\label{7bb}
\eeqn
In general $h_3, h_4, h_5, h_3', h_4',h_5',  h_3'', h_4'',h_5'', h_6, h_7, h_8$ can be complex and we define their phases
so that

\beqn
h_k= |h_k| e^{i\chi_k}, ~~h_k'= |h_k'| e^{i\chi_k'}, ~~~h_k''= |h_k''| e^{i\chi_k''}\,.
\label{mix}
\eeqn
The squark sector of the model contains a variety of terms including F -type, D-type and SUSY soft breaking terms.   The details of these contributions to squark mass
square matrices are  discussed in {section} \ref{A}.

{
\section{The analysis of Chromoelectric Dipole Moment Operator \label{sec3} }
 The chromoelectric dipole moment $\tilde{d}^C$ is the coefficient of the  effective dimension $5$ operator
 which is defined by
\beq
{\cal{L}}_I=-\frac{i}{2} \tilde{d}_q^C \bar{q} \sigma_{\mu \nu}\gamma_5 T^a q G^{\mu \nu a},
\label{reduced}
\eeq
where  $G^{\mu\nu a}$ is the gluon field strength and $T^a$ are the $SU(3)$ generators.
The   quarks will have five different contributions to the CEDM arising from the W, Z, gluino, chargino and neutralino
 exchanges. We denote these contributions  by $\tilde{d^C_u}(W)$, $\tilde{d^C_u}(Z)$, $\tilde{d^C_u}({\tilde{g}})$, $\tilde{d^C_u}({\chi^+})$ and $\tilde{d^C_u}({\chi^0})$.  We discuss each of these contributions below.

\subsection{ W  exchange contribution to quark CEDM \label{sec3.1}}
For the up quark
the W- exchange contribution arises from  the left diagram of Fig.~(1) using the interaction of Eq. (13), i.e.,

 \begin{figure}
\begin{center}
      \includegraphics[scale=.5]{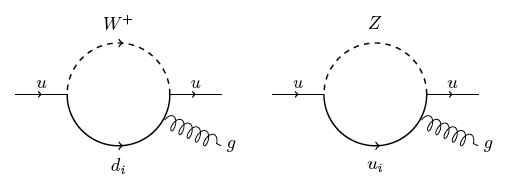}
          \caption{$W$ and  $Z$ exchange contributions to the CEDM of the up quark.
          Similar exchange contributions exist for the CEDM of the down quark where
          $u$ and $d$ are interchanged and $W^+$ is replaced by $W^-$ in the diagrams above.
        }
\end{center}
\label{fig1}
\end{figure}

\begin{align}
-{\cal L}_{d W u} &= W^{\dagger}_{\rho}\sum_{i=1}^{5}\sum_{j=1}^{5}\bar{u}_{j}\gamma^{\rho}[G_{L_{ji}}^W P_L + G_{R_{ji}}^W P_R]d_{i}+\text{h.c.},
\label{w-int}
\end{align}
where $G^W_L$ and $G^W_R$ are defined in section 8. The contribution of the W-exchange graph to $\tilde d^C_u$ is given by

\begin{align}
\tilde{d^C_u}(W)&=\frac{g_s}{16\pi^2}\sum_{i=1}^{5}\frac{m_{d_i}}{m^2_W}\text{Im}(G^{W}_{L 4i}G^{W*}_{R 4i })
I_1\left(\frac{m^{2}_{{d}_{i}}}{m^{2}_{W}},
\frac{m^{2}_{{u}_{4}}}{m^{2}_{W}}\right)\,,
\label{W-contribution}
\end{align}
where  $I_1( r_1, r_2)$  is a form factor given by

\begin{align}
I_1(r_1, r_2)&=\int_{0}^{1} dx \frac{(4 + r_1 -r_2) x - 4 x^2}{1+ (r_1 -r_2 -1)x +r_2 x^2}\,.
\end{align}

In the limit when $r_2$ is very small as the case here, this integral gives the closed form
\begin{align}
I_1(r_1,0) = \frac{2}{(1-r_1)^{2}}\left[1+\frac{1}{4}r_1 +\frac{1}{4}r_1^2+\frac{3 r_1\ln r_1}{2(1-r_1)} \right].
\end{align}

The W contribution to the down quark CEDM is given by
\begin{align}
\tilde{d^C_d}(W)&=\frac{g_s}{16\pi^2}\sum_{i=1}^{5}\frac{m_{u_i}}{m^2_W}\text{Im}(G^{W*}_{L i4}G^{W}_{R i4 })
I_1\left(\frac{m^{2}_{{u}_{i}}}{m^{2}_{W}},
\frac{m^{2}_{{d}_{4}}}{m^{2}_{W}}\right)\,.
\end{align}

\subsection{Z exchange contribution to quark CEDM \label{sec3.2}}

For the Z boson exchange the interactions that enter with the up type quarks are given by

\begin{align}
-{\cal L}_{uu Z} &= Z_{\rho}\sum_{j=1}^{5}\sum_{i=1}^{5}\bar{u}_{j}\gamma^{\rho}[C_{L_{ji}}^{uZ} P_L + C_{R_{ji}}^{uZ} P_R]u_{i},
\label{z-u-int}
\end{align}
where the couplings $C_L^{uZ}$ and $C_R^{uZ}$ are defined in section 8.
Using this interaction the computation of  the Z exchange contributions to the up quarks is given by the loop diagram to the right in Fig. (1). Its contribution is

\begin{align}
\tilde{d^C_u}(Z)&=\frac{g_s}{16\pi^2}\sum_{i=1}^{5}\frac{m_{u_i}}{m^2_Z}\text{Im}(C_{L_{4i}}^{uZ} C_{R_{4i}}^{uZ*})
I_1\left(\frac{m^{2}_{{u}_{i}}}{m^{2}_{Z}}
,\frac{m^{2}_{{u}_{4}}}{m^{2}_{Z}}
\right)\,.
\end{align}

For the Z boson exchange, the interactions that enter with the down type quarks are given by

\begin{align}
-{\cal L}_{dd Z} &= Z_{\rho}\sum_{j=1}^{5}\sum_{i=1}^{5}\bar{d}_{j}\gamma^{\rho}[C_{L_{ji}}^{dZ} P_L + C_{R_{ji}}^{dZ} P_R]d_{i},
\label{z-d-int}
\end{align}
where the couplings $C_L^{dZ}$ and $C_R^{dZ}$ are as defined in section 8.
A calculation similar to that of the up quark CDEM gives  a contribution to the d-quark moment  so that

\begin{align}
\tilde{d^C_d}(Z)&=\frac{g_s}{16\pi^2}\sum_{i=1}^{5}\frac{m_{d_i}}{m^2_Z}\text{Im}(C_{L_{4i}}^{dZ} C_{R_{4i}}^{dZ*})
I_1\left(\frac{m^{2}_{{d}_{i}}}{m^{2}_{Z}}
,\frac{m^{2}_{{d}_{4}}}{m^{2}_{Z}}
\right)\,.
\end{align}

\subsection{Chargino exchange contribution to CEDM\label{sec3.3}}

 In this section we discuss the  interactions in the mass diagonal basis involving squarks,
 charginos and quarks.  Thus we have
\begin{align}
-{\cal L}_{u-\tilde{d}-\chi^{-}} &= \sum_{j=1}^{5}\sum_{i=1}^{2}\sum_{k=1}^{10}\bar{u}_{j}(C_{jik}^{Lu}P_{L}+C_{jik}^{Ru}P_{R})\tilde{\chi}^{ci}\tilde{d}_{k}+\text{h.c.},
\label{d-chargino}
\end{align}
and

\begin{align}
-{\cal L}_{d-\tilde{u}-\chi^{-}} &= \sum_{j=1}^{5}\sum_{i=1}^{2}\sum_{k=1}^{10}\bar{d}_{j}(C_{jik}^{Ld}P_{L}+C_{jik}^{Rd}P_{R})\tilde{\chi}^{ci}\tilde{u}_{k}+\text{h.c.},
\label{u-chargino}
\end{align}
where the couplings $C^{Lu}$, $C^{Ru}$, $C^{Ld}$ and $C^{Rd}$ and  are as defined in section 8.
The loop contributions to the up -quark CEDM arise from the right diagram of Fig.~(2). Their contribution to CEDM of quarks using
 Eq.~(22) and Eq.~(23) are given by

\begin{figure}
\begin{center}
               \includegraphics[scale=.6]{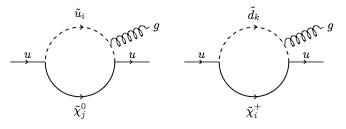}
                \caption{Left diagram:
                     Supersymmetric loop contributions to the CEDM of the up-quark from the diagram involving the
                exchange of neutralinos and up-squarks.
   Right diagram:
        Chargino  and          down-squark loop contribution to the CEDM of the up quark.
Similar loop contributions exist
          for the CEDM of the down quark, where $u$ and $d$ are interchanged, $\tilde u$ and $\tilde d$ are
          interchanged and $\chi^+$ is replaced by $\chi^-$ in the diagrams above.
           }
\end{center}
\label{fig2}
\end{figure}

\begin{align}
\tilde{d^C_u}({\chi^+})&=\frac{g_s}{16\pi^2}\sum_{i=1}^{2}\sum_{k=1}^{10}\frac{m_{\chi^{+}_i}}{M^2_{\tilde{d}_{k}}}\text{Im}(C^{Lu}_{4ik}C^{Ru*}_{4ik})
I_3\left(\frac{m^{2}_{\chi^{+}_i}}{M^{2}_{\tilde{d}_{k}}}
,\frac{m^{2}_{{u}_{4}}}{M^{2}_{\tilde{d}_{k}}}
\right)\,,
\label{3.1u}
\end{align}

\begin{align}
\tilde{d^C_d}({\chi^+})&=\frac{g_s}{16\pi^2}\sum_{i=1}^{2}\sum_{k=1}^{10}\frac{m_{\chi^{+}_i}}{M^2_{\tilde{u}_{k}}}\text{Im}(C^{Ld}_{4ik}C^{Rd*}_{4ik})
I_3\left(\frac{m^{2}_{\chi^{+}_i}}{M^{2}_{\tilde{u}_{k}}}
,\frac{m^{2}_{{d}_{4}}}{M^{2}_{\tilde{u}_{k}}}
\right),
\label{3.1d}
\end{align}
where $I_3(r_1, r_2)$ is given by
\begin{equation}
I_3(r_1, r_2) = \int_{0}^{1} dx \frac{x - x^2}{1+(r_1 -r_2 -1)x +r_2 x^2}\,.
\end{equation}
In the limit when $r_2$ is very small as is the case here we have the closed form
\begin{equation}
I_3(r_1,0)=  \frac{1}{2(r_1-1)^2} \left(1+r_1 + \frac{2r_1 \ln r_1}{1-r_1}\right).
\label{2.2}
\end{equation}

\subsection{Neutralino exchange contribution to CEDM\label{sec3.4}}
 We now  discuss the  interactions in the mass diagonal basis involving up quarks,
 up squarks and neutralinos.  Thus we have,

\begin{align}
-{\cal L}_{u-\tilde{u}-\chi^{0}} &= \sum_{i=1}^{5}\sum_{j=1}^{4}\sum_{k=1}^{10}\bar{u}_{i}(C_{uijk}^{'L}P_{L}+C_{uijk}^{'R}P_{R})\tilde{\chi}^{0}_{j}\tilde{u}_{k}+\text{h.c.},
\label{u-neutralino}
\end{align}
The interaction of the down quarks, down squarks and neutralinos is given by

\begin{align}
-{\cal L}_{d-\tilde{d}-\chi^{0}} &= \sum_{i=1}^{4}\sum_{j=1}^{4}\sum_{k=1}^{10}\bar{d}_{i}(C_{dijk}^{'L}P_{L}+C_{dijk}^{'R}P_{R})\tilde{\chi}^{0}_{j}\tilde{d}_{k}+\text{h.c.},
\label{d-neutralino}
\end{align}
where the couplings $C^{'L}$ and $C^{'R}$ as given in section 8.
Using the interactions of Eq.~(28) the neutralino exchange contribution to the CEDM of the up-quark  is given by

\begin{align}
\tilde{d^C_u}({\chi^0})&=\frac{g_s}{16\pi^2}\sum_{i=1}^{4}\sum_{k=1}^{10}\frac{m_{\chi^{0}_i}}{M^2_{\tilde{u}_{k}}}\text{Im}(C^{'L}_{u4ik}C^{'R*}_{u4ik})
 I_3\left(\frac{m^{2}_{\chi^{0}_i}}{M^{2}_{\tilde{u}_{k}}}
,\frac{m^{2}_{{u}_{4}}}{M^{2}_{\tilde{u}_{k}}}
\right)\,.
\label{3.1n1}
\end{align}
Similarly using the interactions  of  Eq.~(29)  the CEDM of the down quark is given by

\begin{align}
\tilde{d^C_d}({\chi^0})&=\frac{g_s}{16\pi^2}\sum_{i=1}^{4}\sum_{k=1}^{10}\frac{m_{\chi^{0}_i}}{M^2_{\tilde{d}_{k}}}\text{Im}(C^{'L}_{d4ik}C^{'R*}_{d4ik})
 I_3\left(\frac{m^{2}_{\chi^{0}_i}}{M^{2}_{\tilde{d}_{k}}}
,\frac{m^{2}_{{d}_{4}}}{M^{2}_{\tilde{d}_{k}}}
\right),
\label{3.1n2}
\end{align}

\subsection{Gluino exchange contribution to CEDM  \label{sec3.5}}

\begin{align}
-{\cal L}_{qq \tilde{g}} &=\sqrt{2} g_s  \sum_{j=1}^{3}\sum_{k=1}^{3}\sum_{a=1}^{8}\sum_{l=1}^{5}
\sum_{m=1}^{10}\
T^a_{jk}
\bar{q}^j_{l}[C_{L_{lm}} P_L + C_{R_{lm}}P_R]
\tilde{g}_a
\tilde{q}^k_{m} + \text{h.c.},
\label{q-gluino}
\end{align}
where the couplings $C_{L_{lm}}$ and  $C_{R_{lm}}$ are defined in section 8. Using Eq.~(32) the gluino exchange contribution to the up  quark CEDM arising from the loop diagrams of Fig.~3 is given by

\begin{figure}
\begin{center}
      \includegraphics[scale=.6]{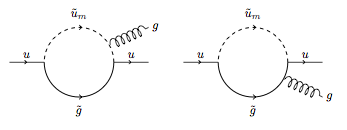}
                \caption{
           Left diagram:
                Supersymmetric loop contributions to the CEDM of the up-quark arising from the  exchange of gluino and up squarks with the gluon emission
 from the internal up squark line.  Right diagram: Same as left diagram except that the gluon emission is from
 the internal gluino line.  Similar loop contributions exist
          for the CEDM of the down quark, where $u$ and $d$ are interchanged, $\tilde u$ and $\tilde d$ are
          interchanged.
           }
\end{center}
\label{fig3}
\end{figure}
\begin{align}
\tilde{d^C_u}({\tilde{g}})&=\frac{g_s \alpha_s}{12\pi^2}\sum_{m=1}^{10}\frac{m_{\tilde{g}}}{M^2_{\tilde{u}_m}}\text{Im}(K_{L_{um}} K^*_{R_{um}})
I_5\left(\frac{m^{2}_{\tilde{g}}}{M^{2}_{\tilde{u}_m}}
,\frac{m^{2}_{u4}}{M^{2}_{\tilde{u}_m}}
\right)\,.
\end{align}
Similarly using  Eq.~(32) the  gluino contribution to the down quark CEDM is given by

\begin{align}
\tilde{d^C_d}({\tilde{g}})&=\frac{g_s \alpha_s}{12\pi^2}\sum_{m=1}^{10}\frac{m_{\tilde{g}}}{M^2_{\tilde{d}_m}}\text{Im}(K_{L_{dm}} K^*_{R_{dm}})
I_5\left(\frac{m^{2}_{\tilde{g}}}{M^{2}_{\tilde{d}_m}}
,\frac{m^{2}_{d4}}{M^{2}_{\tilde{d}_m}}
\right)\,.
\end{align}
Here $K_{L_{qm}}$ and $K_{R_{qm}}$ are given by

\begin{align}
K_{L_{qm}}=(D^{q*}_{R24} \tilde{D}^{q}_{4m}
-D^{q*}_{R54} \tilde{D}^{q}_{10m}
-D^{q*}_{R44} \tilde{D}^{q}_{8m}-
D^{q*}_{R34} \tilde{D}^{q}_{6m}
-D^{q*}_{R14} \tilde{D}^{q}_{3m}
)e^{-i\xi_3/2}\,,
\label{KL}
\end{align}
and
\begin{align}
K_{R_{qm}}=(D^{q*}_{L44} \tilde{D}^{q}_{7m}
+D^{q*}_{L54} \tilde{D}^{q}_{9m}
+D^{q*}_{L34} \tilde{D}^{q}_{5m}+
D^{q*}_{L14} \tilde{D}^{q}_{1m}
-D^{q*}_{L24} \tilde{D}^{q}_{2m}
)e^{i\xi_3/2}\,,
\label{KR}
\end{align}
where $I_5(r_1, r_2)$ is the loop function defined by
\begin{equation}
I_5(r_1, r_2) = \int_{0}^{1} dx \frac{x+8x^2}{1+(r_1 - r_2 -1) x +r_2 x^2}\,.
\end{equation}
In the limit where $r_2$ is very small as is the case here we  get the closed form
\begin{equation}
 I_5(r_1,0)=\frac{1}{2(r_1-1)^2} \left(10r_1-26 + \frac{2 r_1 \ln r_1}{1-r_1}-\frac{18\ln r_1}{1-r_1}\right).
\label{2.6}
\end{equation}

\section{The neutron CEDM  \label{sec4}}
As discussed in the previous section, the total contribution to CEDM of the quarks consists of
five contributions arising from the exchange of the W, the Z, the charginos, the neutralinos, and the gluino,
so that

\beq
\tilde d^C_q=
 \tilde{d^C_q}(W) + \tilde{d^C_q}(Z)   +\tilde{d^C_q}({\chi^+})+ \tilde{d^C_q}({\chi^0})
 + \tilde{d^C_q}({\tilde{g}}),~~q=u,d\,.
\eeq
The contribution of the chromoelectric  operator to the EDMs of quarks can be computed using  dimensional analysis~\cite{Manohar:1983md}.
 The contribution to the quark EDM arising from $\tilde d^C_q$ is given by
\begin{equation}
d_q^C = \frac{e}{4 \pi} \eta^C \tilde{d_q^C}\,,
\end{equation}
where $\eta^C$ is approximately equal to $3.4$.  The factor $\eta^C$
  brings the electric dipole moment from the electroweak scale down to the hadronic scale
where it can be compared with experiment.
To obtain the contribution to  the neutron EDM from the quark EDM, we use the non-relativistic $SU(6)$  quark model which gives
\begin{equation}
d^C_n= \frac{1}{3} [4 d^C_d- d^C_u]\,.
\label{7.1}
\end{equation}

\section{Numerical analysis of neutron EDM\label{sec5}}
The current experimental limit on the EDM of the
neutron is~\cite{Baker:2006ts}
\beq
|d_n|<  2.9 \times 10^{-26} ~e{\rm cm} ~~~(90\% ~{\rm CL}).
\label{edm}
\eeq
It is expected that a higher sensitivity by as much as two orders of magnitude more sensitive than the current limit
may be achievable in the future~\cite{Ito:2007xd}. \\

We present now a numerical analysis of the neutron CEDM
 first for the case of MSSM and next for the MSSM extension. The first analysis involves no mixing with the mirror generation and {with}  the fourth sequential generation and the only CP phases that appear are those from the MSSM sector. Thus in this case all the mixing parameters, given in Eq.~(11), are set to zero.
 The  second analysis is for the MSSM extension where the mixings of the mirror generation and  of the fourth sequential generation with the three generations are switched on. The results are given in Table~\ref{table:2} and Figs.~\ref{fig444}-\ref{fig9}. In the analysis, in the squark sector we assume $m^{u^2}_0=M^2_{\tilde T}=M^2_{\tilde t_1}=M^2_{\tilde t_2}=M^2_{\tilde t_3}$ and $m^{d^2}_0=M^2_{\tilde 1 L}=M^2_{\tilde B}=M^2_{\tilde b_1}=M^2_{\tilde Q}=M^2_{\tilde 2 L}=M^2_{\tilde b_2}=M^2_{\tilde 3 L}=M^2_{\tilde b_3}$. {{To simplify the numerical analysis further we assume}} $m^u_0=m^d_0=m_0$.  Additionally the trilinear couplings are chosen {so that}: $A^u_0=A_t=A_T=A_c=A_u=A_{4t}$ and $A^d_0=A_b=A_B=A_s=A_d=A_{4b}$.
 { The input parameters are such that the sparticle spectrum that enters  the loop are consistent with
the current experimental limits  from the LHC in each of the cases, i.e., with or without mixing.}
  \\

\begin{table}[H]
\begin{center}
\begin{tabular}{l  c  c  c  c }
\hline\hline
   & \multicolumn{2}{c}{(i)} & \multicolumn{2}{c}{(ii)} \\
\cline{2-5}
Contribution & Up & Down & Up & Down \\
\hline
Chargino, $d^{\chi^{\pm}}_q$ & $2.49\times10^{-29}$ & $-1.29\times10^{-26}$ & $2.16\times10^{-29}$ & $-2.08\times10^{-26}$ \\
Neutralino, $d^{\chi^0}_q$ & $-2.49\times10^{-32}$ & $4.75\times10^{-29}$ & $-2.90\times10^{-32}$ & $5.47\times10^{-29}$ \\
Gluino, $d^g_q$ & $3.42\times10^{-29}$ & $-4.24\times10^{-28}$ & $7.49\times10^{-28}$ & $2.06\times10^{-26}$\\
Total, $d_q$ & $5.90\times10^{-29}$ & $-1.32\times10^{-26}$ & $7.71\times10^{-28}$ & $-1.42\times10^{-28}$ \\
EDM, $d^E_n$ & \multicolumn{2}{c}{$-2.70\times10^{-26}$}  &  \multicolumn{2}{c}{$-6.83\times10^{-28}$}\\
\hline
Chargino, $d^{C}_q(\chi^{\pm})$ & $-3.41\times10^{-30}$ & $-2.15\times10^{-27}$ & $-2.89\times10^{-30}$ & $-3.40\times10^{-27}$ \\
Neutralino, $d^{C}_q(\chi^0)$ & $-4.54\times10^{-32}$ & $-1.73\times10^{-28}$ & $-5.30\times10^{-32}$ & $-2.00\times10^{-28}$ \\
Gluino, $d^C_q(\tilde{g})$ & $5.51\times10^{-29}$ & $1.37\times10^{-27}$ & $1.21\times10^{-27}$ & $-6.63\times10^{-26} $\\
Total, $d^{C}_q$ & $1.40\times10^{-29}$ & $-2.58\times10^{-28}$ & $3.26\times10^{-28} $ & $-1.89\times10^{-26}$ \\
CEDM, $d^C_n$ & \multicolumn{2}{c}{$-3.49\times10^{-28}$}  &  \multicolumn{2}{c}{$-2.53\times10^{-26}$} \\
\hline \hline
\end{tabular}
\caption{An exhibition of the chargino, neutralino and gluino exchange contributions to the quark and the neutron EDM, CEDM and their sum for the case when there is no mixing of the vectorlike generation with the three generations. The analysis is for two benchmark points (i) and (ii). Benchmark (i): $\theta_{\mu}=3.3\times10^{-3}$, $\xi_3=1\times10^{-3}$. Benchmark (ii): $\theta_{\mu}=4.7\times10^{-3}$, $\xi_3 = 3.6$. The common parameters are: $\tan\beta = 40$, $m_0=m^u_0=m^d_0=3000$, $|m_1| = 185$, $|m_2| = 220$, $|A^u_0| = 680$, $|A^d_0| = 600$, $|\mu| = 400$, $m_g = 1000$, $|h_3| = |h'_3| = |h''_3| = |h_4| = |h'_4| = |h''_4| = |h_5| = |h'_5| = |h''_5| = |h_6| = |h_7| =|h_8| = 0$, $\xi_1 = 2\times10^{-2}$, $\xi_2 = 2\times10^{-3}$, $\alpha_{A^u_0} = 2\times10^{-2}$, $\alpha_{A^d_0} = 3$. All masses are in GeV, all phases in rad and the electric dipole moment in $e$cm.}
\label{table:1}
\end{center}
\end{table}

{We discuss now in further detail the cases without and with mixing with the vectorlike generation.
We begin with the case with no mixing.}
In table~\ref{table:1}, we give the individual contributions to the up and down quark EDM and CEDM, namely, the chargino,
the neutralino and the gluino contributions. The W and Z contributions are not shown since they are
{absent}
 in this case of no mixing with the vectorlike generation and the fourth sequential generation. The different contributions are given for two benchmark points (i) and (ii), where in the first, the neutron EDM dominates the neutron CEDM and in the second, the opposite is {the case}. The chargino and gluino contributions are the main contributors, while the neutralino contribution is suppressed. Note that the total neutron EDM, $|d_n|$, obtained by adding $d^E_n$ and $d^C_n$ in the table satisfy Eq.~(42). Another observation is the largeness of the down quark contribution in comparison with its up quark counterpart. This is attributed to the large value of $\tan\beta$ which tends to enhance the down quark couplings.

\begin{table}[H]
\begin{center}
\begin{tabular}{l  c  c  c  c }
\hline\hline
   & \multicolumn{2}{c}{(i)} & \multicolumn{2}{c}{(ii)} \\
\cline{2-5}
Contribution & Up & Down & Up & Down \\
\hline
Chargino, $d^{\chi^{\pm}}_q$ & $7.65\times10^{-30}$ & $-6.91\times10^{-27}$ & $7.08\times10^{-30}$ & $-8.27\times10^{-27}$ \\
Neutralino, $d^{\chi^0}_q$ & $3.93\times10^{-32}$ & $6.90\times10^{-30}$ & $3.91\times10^{-32}$ & $7.32\times10^{-30}$ \\
Gluino, $d^g_q$ & $-2.01\times10^{-28}$ & $-5.35\times10^{-27}$ & $2.25\times10^{-28}$ & $5.91\times10^{-27}$\\
W Boson, $d^W_q$ & $3.77\times10^{-30}$ & $3.46\times10^{-28}$ & $3.77\times10^{-30}$ & $3.46\times10^{-28}$ \\
Z Boson, $d^Z_q$ & $8.02\times10^{-31}$ & $3.05\times10^{-29}$ & $8.02\times10^{-31}$ & $3.05\times10^{-29}$\\
Total, $d_q$ & $-1.88\times10^{-28}$ & $-1.19\times10^{-26}$ & $2.37\times10^{-28}$ & $-1.97\times10^{-27}$ \\
EDM, $d^E_n$ & \multicolumn{2}{c}{$-2.41\times10^{-26}$}  &  \multicolumn{2}{c}{$-4.13\times10^{-27}$}\\
\hline
Chargino, $d^{C}_q(\chi^{\pm})$ & $-7.66\times10^{-31}$ & $-8.98\times10^{-28}$ & $-6.95\times10^{-31}$ & $-1.07\times10^{-27}$ \\
Neutralino, $d^{C}_q(\chi^0)$ & $7.17\times10^{-32}$ & $-2.52\times10^{-29}$ & $7.14\times10^{-32}$ & $-2.68\times10^{-29}$ \\
Gluino, $d^C_q(\tilde{g})$ & $-4.57\times10^{-28}$ & $2.44\times10^{-26}$ & $5.13\times10^{-28}$ & $-2.70\times10^{-26} $\\
W Boson, $d^C_q(W)$ & $-2.97\times10^{-30}$ &  $2.29\times10^{-28}$ & $-2.97\times10^{-30}$ & $2.29\times10^{-28}$ \\
Z Boson, $d^C_q(Z)$ & $1.46\times10^{-30}$ & $-1.11\times10^{-28}$ & $1.46\times10^{-30}$ & $-1.11\times10^{-28}$ \\
Total, $d^C_q$ & $-1.24\times10^{-28}$ & $6.38\times10^{-27}$ & $1.38\times10^{-28} $ & $-7.56\times10^{-27}$ \\
CEDM, $d^C_n$ & \multicolumn{2}{c}{$8.54\times10^{-27}$}  &  \multicolumn{2}{c}{$-1.01\times10^{-26}$} \\
\hline \hline
\end{tabular}
\caption{An exhibition of the chargino, neutralino, gluino, $W$ and $Z$ exchange contributions to the quark and the neutron EDM, CEDM and their sum for the case when there is mixing of the vectorlike generation with the three generations. The analysis is for two benchmark points (i) and (ii). Benchmark (i): $\theta_{\mu}=4\times10^{-3}$, $\xi_3 =1.12$. Benchmark (ii): $\theta_{\mu} =4.6\times10^{-3}$, $\xi_3 = 4.71$. The common parameters are: $\tan\beta = 40$, $m_0=m^u_0=m^d_0=5500$, $|m_1| = 185$, $|m_2| = 220$, $|A^u_0| = 680$, $|A^d_0| = 600$, $|\mu| = 400$, $m_g = 1100$, $m_T = 300$, $m_B = 240$, $m_{4t} = 320$, $m_{4b} = 280$, $|h_3| = 1.58$, $|h'_3| = 6.34\times10^{-2}$, $|h''_3| = 1.97\times10^{-2}$, $|h_4| = 4.42$, $|h'_4| = 5.07$, $|h''_4| = 12.87$, $|h_5| = 6.6$, $|h'_5| = 2.67$, $|h''_5| = 1.86\times10^{-1}$, $|h_6| = 1000$, $|h_7| = 1000$, $|h_8| = 1000$, $\xi_1 = 2\times10^{-2}$, $\xi_2 = 2\times10^{-3}$, $\alpha_{A^u_0} = 2\times10^{-2}$, $\alpha_{A^d_0} = 3$, $\chi_3 = 2\times10^{-2}$, $\chi'_3 = 1\times10^{-3}$, $\chi''_3 = 4\times10^{-3}$, $\chi_4 = 7\times10^{-3}$, $\chi'_4 = \chi''_4 = 1\times10^{-3}$, $\chi_5 = 9\times10^{-3}$, $\chi'_5 = 5\times10^{-3}$, $\chi''_5 = 2\times10^{-3}$, $\chi_6 =\chi_7 =\chi_8 = 5\times10^{-3}$. All masses are in GeV, all phases in rad and the electric dipole moment in $e$cm.}
\label{table:2}
\end{center}
\end{table}

{Next we consider the case with mixings. The results are presented in
 table~\ref{table:2} for two benchmark points (i) and (ii).  Here in addition to the chargino, the neutralino and the gluino exchanges one also has $W$ and $Z$ exchanges. The analysis shows the  dominance of the EDM over the CEDM for benchmark (i) while opposite is the case for benchmark
(ii).  The total EDM for each benchmark point satisfies the experimental constraints of
 Eq.~(42).  Here we note that the EDM and CEDM are constrained not only by the experimental
 limits on the MSSM spectrum, but also by the limits on new quarks.  Thus for the benchmarks presented in Table 2,
 the vectorlike quarks have masses gotten by diagonalization of the
matrices of  Eq.~(5) and Eq.~(10) and  are given in {Table}~\ref{table:3}. The
results of Table \ref{table:3} are consistent with~\cite{pdg}.
More stringent constraints on these masses  will be available at the  LHC RUN-II.}

\begin{table}[H]
\begin{center}
\begin{tabular}{l l}
   \hline
   Mirror Up Quark  & $m_{t'}$ = 1037 GeV  \\
  Mirror Down Quark  & $m_{b'}$ = 740 GeV  \\
  Fourth G Up Quark & $m^{\rm up}_{4}$ = 1057 GeV \\
  Fourth G Down Quark & $m^{\rm down}_{4}$ = 1260 GeV \\
  \hline\hline
\end{tabular}
\caption{{An exhibition of the masses of  the heavy extra quarks corresponding to the parameter space of table~\ref{table:2}.}}
\label{table:3}
\end{center}
\end{table}


\begin{figure}[H]
\begin{center}
{\rotatebox{0}{\resizebox*{10cm}{!}{\includegraphics{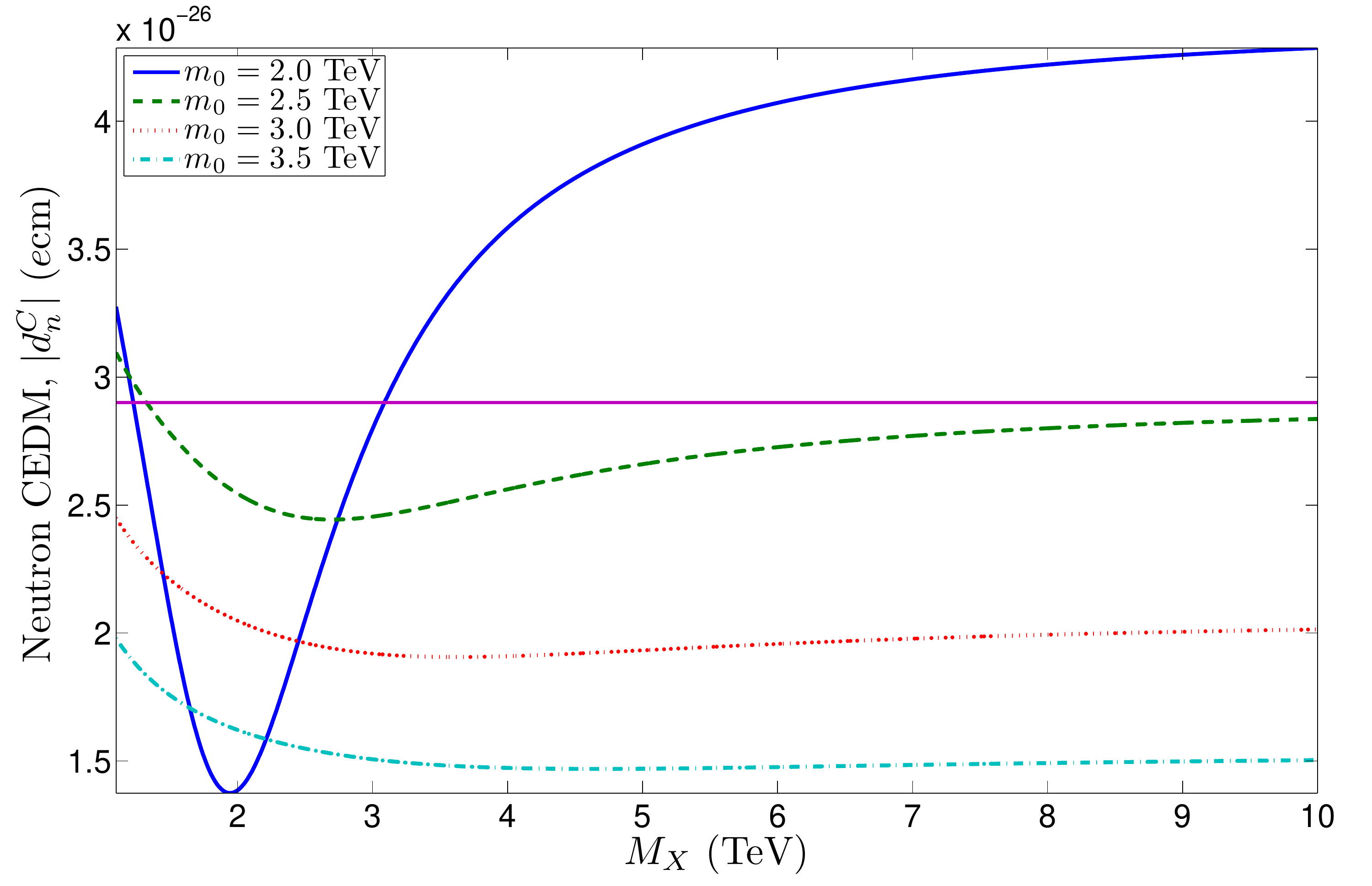}}\hglue5mm}}
\caption{Variation of neutron CEDM $|d^C_{n}|$ versus $M_X$ $(M_X=|h_6|=|h_7|=|h_8|)$, for four values of $m_0$. From top to bottom at $M_X=4$ TeV, $m_0=m^u_0=m^d_0$ = 2.0, 2.5, 3.0, 3.5 TeV. Other parameters have the values $\tan\beta=14$, $|m_{1}|=185$, $|m_{2}|=220$, $|\mu|=350$, $|A^{u}_{0}|=680$, $|A^{d}_{0}|=600$, $m_{T}=300$, $m_{B}=260$, $m_g=1000$, $m_{4t}=320$, $m_{4b}=280$, $|h_{3}|=1.58$, $|h'_{3}|=|h''_{3}|=RM_X$, $|h_{4}|=4.42$, $|h'_{4}|=|h''_{4}|=RM_X$, $|h_{5}|=6.6$, $|h'_{5}|=|h''_{5}|=RM_X$, $R=1\times 10^{-3}$, $\theta_{\mu}=3.98$, $\xi_{1}=\xi_{2}=4.52$, $\xi_3=2.42$, $\alpha_{A^{u}_{0}}=5.0$, $\alpha_{A^{d}_{0}}=1.14$, $\chi_{3}=2.38$, $\chi'_{3}=4.92$, $\chi''_{3}=2.58$, $\chi_{4}=4.86$, $\chi'_{4}=1.6$, $\chi''_{4}=1.37$, $\chi_{5}=1.14$, $\chi'_{5}=4.39$, $\chi''_{5}=2.38$, $\chi_{6}=4.92$, $\chi_{7}=2.58$, $\chi_{8}=4.86$. All masses are in GeV and phases in rad.}
\label{fig444}
\end{center}
\end{figure}
{
Next we give an analysis of the quark CEDMs dependence on the mass scales as well as on the CP phases both in the MSSM sector
as well as the new sector. Thus the CEDM depends on the mass scale of the vectorlike sector and in the MSSM sector
it depends on the universal scalar mass $m_0$, and on the gaugino mass scales. Further, it has dependence on
several CP phases both from the MSSM sector as well as from the vectorlike sector. We discuss the dependence of the
CEDM on the mass scales first, and specifically on the mass scales $M_X$ (from the vectorlike sector) and on $m_0$ and
on $m_g$. }

{ Fig.~\ref{fig444} gives  the dependence of the effect of the vectorlike generation on
CEDM where we exhibit CEDM vs}
$M_X$, where $M_X=|h_6|=|h_7|=|h_8|$ and that $|h'_3|=|h''_3|=|h'_4|=|h''_4|=|h'_5|=|h''_5|=RM_X$
{while $R=1\times 10^{-3}$. We note that the allowed range of values for $R$ is highly constrained.
Thus smaller values of $R$ will not produce interesting results while larger values are likely to upset the
quark masses  for the first three generations.}
 The analysis shows that  CEDM lower than the upper limit can be obtained and masses in the TeV range may be probed using the constraint given by Eq.~(42) which should undergo further refinements in the future. The curve corresponding to $m_0=2$ TeV is characterized by a dip at $M_X\sim 1.9$ TeV. This dip quickly widens and is replaced by a shallow drop for $m_0=2.5$ TeV and then disappears completely for larger values of $m_0$. The variation of the CEDM eventually levels off for higher values of $M_X$ and $m_0$. Further analysis shows that the dip is caused by a sudden drop in the mass of the lightest up squark mass for $M_X\sim 1.9$ TeV in this region of the parameter space. {The analysis of the dip is rather involved but arises as} a result of the competition among the different components of the chromoelectric dipole moment operators, i.e, the W, Z, chargino, neutralino and gluino contributions. The analysis of Fig.~\ref{fig444} makes clear the very sensitive dependence of the CEDM on the vectorlike mass scale
 and exploration of this dependence is one of the primary motivations of this work.

\begin{figure}[H]
\begin{center}
{\rotatebox{0}{\resizebox*{10cm}{!}{\includegraphics{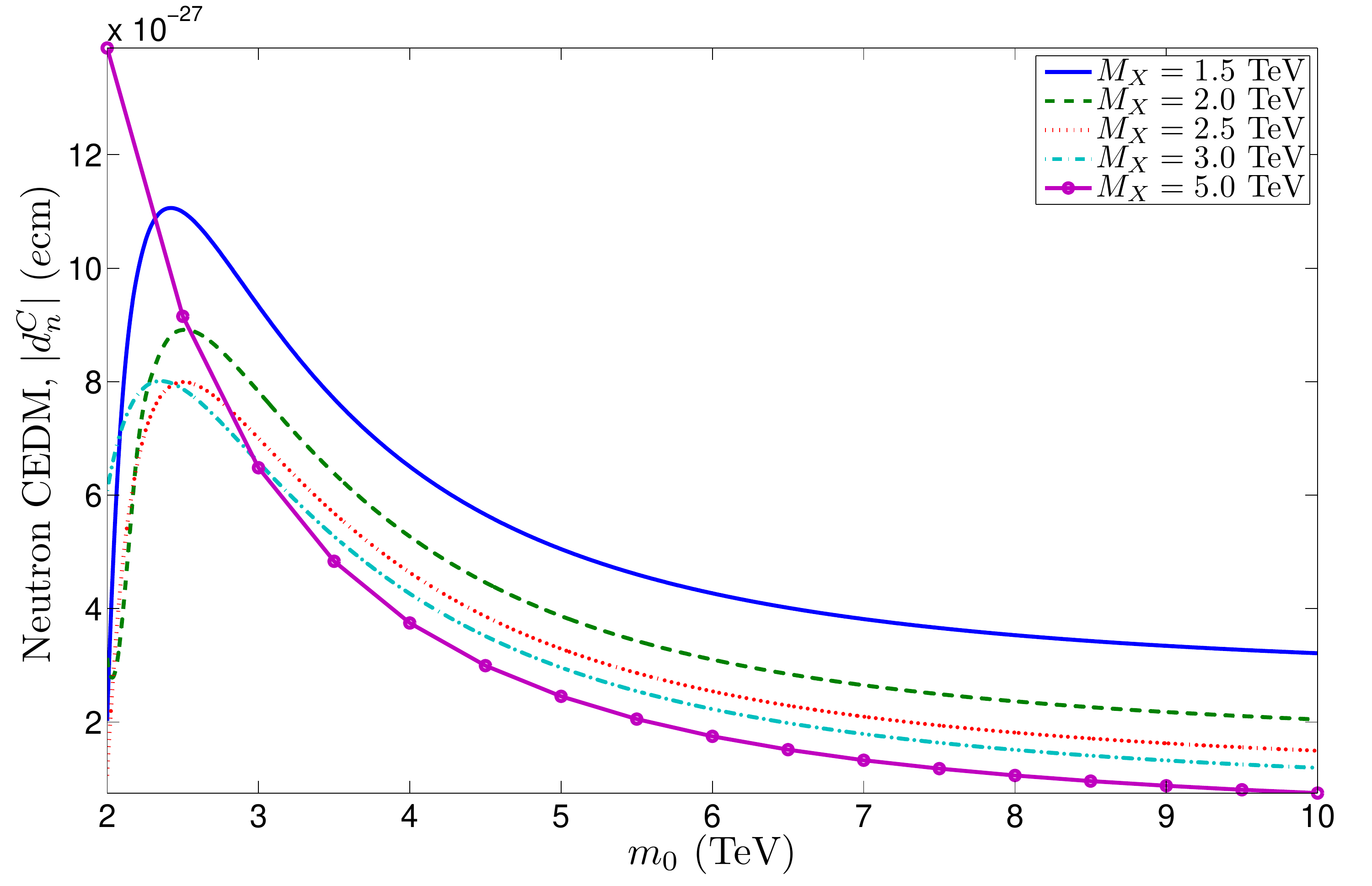}}\hglue5mm}}
\caption{Variation of neutron CEDM $|d^C_{n}|$ versus the scalar mass $m_0$ $(m_0=m^{u}_{0}=m^{d}_{0})$, for five values of $M_X$, $(M_X=|h_6|=|h_7|=|h_8|)$. From top to bottom at $m_0=5$ TeV, $M_X=1.5,2.0,2.5,3.0,5$ TeV. Other parameters have the values $\tan\beta=14$, $|m_{1}|=185$, $|m_{2}|=220$, $|\mu|=350$, $|A^{u}_{0}|=680$, $|A^{d}_{0}|=600$, $m_{T}=300$, $m_{B}=260$, $m_g=1000$, $m_{4t}=320$, $m_{4b}=280$, $|h_{3}|=1.58$, $|h'_{3}|=|h''_{3}|=RM_X$, $|h_{4}|=4.42$, $|h'_{4}|=|h''_{4}|=RM_X$, $|h_{5}|=6.6$, $|h'_{5}|=|h''_{5}|=RM_X$, $R=1\times 10^{-3}$, $\theta_{\mu}=3.8$, $\xi_{1}=\xi_{2}=4.52$, $\xi_3=2.42$, $\alpha_{A^{u}_{0}}=5.0$, $\alpha_{A^{d}_{0}}=1.14$, $\chi_{3}=2.38$, $\chi'_{3}=4.92$, $\chi''_{3}=2.58$, $\chi_{4}=4.86$, $\chi'_{4}=1.6$, $\chi''_{4}=1.37$, $\chi_{5}=1.14$, $\chi'_{5}=4.39$, $\chi''_{5}=2.38$, $\chi_{6}=4.92$, $\chi_{7}=2.58$, $\chi_{8}=4.86$. All masses are in GeV and phases in rad.}
\label{fig666}
\end{center}
\end{figure}

Another way for looking at Fig.~\ref{fig444} is to plot the CEDM against $m_0$ for several values of $M_X$ while
{$R$ is fixed at  $1\times 10^{-3}$ in the same region of parameter space.  This is done  in Fig.~\ref{fig666}.} The plot shows peaks between 2 and 3 TeV and then the CEDM decreases gradually for increasing values of $m_0$. The peak is more pronounced for small values of $M_X$ and disappears for larger values (for $M_X=5$ TeV, here). {The peaks occur in regions where $m_0$ and $M_X$ are
comparable in size as shown in this region.}
 All values of the CEDM obtained in  this region of the parameter space lie below the current upper limit.
\\

{
\begin{figure}[h]
\begin{center}
{\rotatebox{0}{\resizebox*{10cm}{!}{\includegraphics{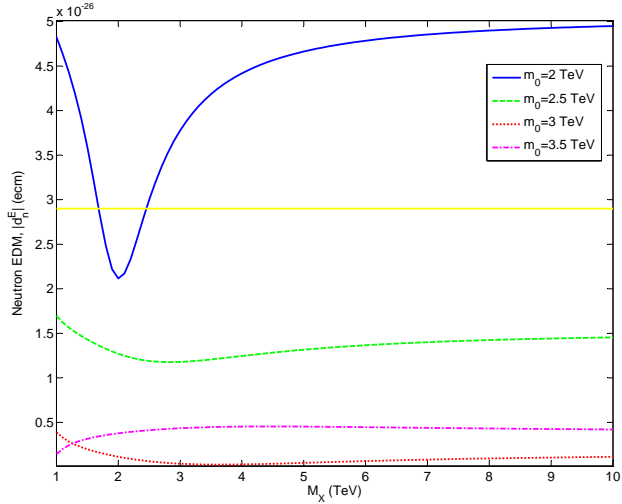}}\hglue5mm}}
\caption{{Variation of neutron EDM $|d^E_{n}|$ versus $M_X$ $(M_X=|h_6|=|h_7|=|h_8|)$, for four values of $m_0$. From top to bottom at $M_X=1$ TeV, $m_0=m^u_0=m^d_0$ = 2.0, 2.5, 3.0, 3.5 TeV. Other parameters have the values $\tan\beta=15$, $|m_{1}|=185$, $|m_{2}|=220$, $|\mu|=350$, $|A^{u}_{0}|=680$, $|A^{d}_{0}|=600$, $m_{T}=300$, $m_{B}=260$, $m_g=1000$, $m_{4t}=320$, $m_{4b}=280$, $|h_{3}|=1.58$, $|h'_{3}|=|h''_{3}|=RM_X$, $|h_{4}|=4.42$, $|h'_{4}|=|h''_{4}|=RM_X$, $|h_{5}|=6.6$, $|h'_{5}|=|h''_{5}|=RM_X$, $R=1\times 10^{-3}$, $\theta_{\mu}=5\times 10^{-3}$, $\xi_{1}=2\times 10^{-2}$, $\xi_{2}=2\times 10^{-3}$, $\xi_{3}=4.0$, $\alpha_{A^{u}_{0}}=2\times 10^{-2}$, $\alpha_{A^{d}_{0}}=3.0$, $\chi_{3}=2\times 10^{-2}$, $\chi'_{3}=1\times 10^{-3}$, $\chi''_{3}=4\times 10^{-3}$, $\chi_{4}=7\times 10^{-3}$, $\chi'_{4}=\chi''_{4}=1\times 10^{-3}$, $\chi_{5}=9\times 10^{-3}$, $\chi'_{5}=5\times 10^{-3}$, $\chi''_{5}=2\times 10^{-3}$, $\chi_{6}=\chi_{7}=\chi_{8}=5\times 10^{-3}$. All masses are in GeV and phases in rad.}}
\label{fig4nn}
\end{center}
\end{figure}
In Fig.~\ref{fig444} we investigated   the dependence of CEDM on the vectorlike mass $M_X$ and found that 
there is a very significant dependence of the CEDM on $M_X$. It is of interest also to examine if the EDM
shows a similar dependence on $M_X$. In Fig.~6 we exhibit this dependence where 
$|d_n^E|$ is plotted against $M_X$ for the same set of $m_0$ values as in  Fig.~\ref{fig444}.
Again as in the case of CEDM we find that EDM  also has a sensitive dependence on $M_X$.
We note here that the analysis of this work for EDM is more general than the analysis of \cite{Aboubrahim:2015nza}.
Thus in ~ \cite{Aboubrahim:2015nza}
we considered only the mixings of the three generations with the mirror generation so that
the quark matrices were $4\times 4$ and the squark square matrices were $8\times 8$ and the 
parameter $M_X$ did not appear in that work. On the other hand, in this work we are considering mixing of the
three generations with the full vectorlike generation consisting of the mirror and the sequential fourth generation.
As a consequence the quark mass matrices are $5\times 5$ and the squark mass squared matrices are $10\times 10$
and this time we have also the dependence on the vectorlike mass $M_X$. Thus the analysis of this work 
is more general than of the work of ~\cite{Aboubrahim:2015nza}.\\
}


\begin{figure}[H]
\begin{center}
{\rotatebox{0}{\resizebox*{7.5cm}{!}{\includegraphics{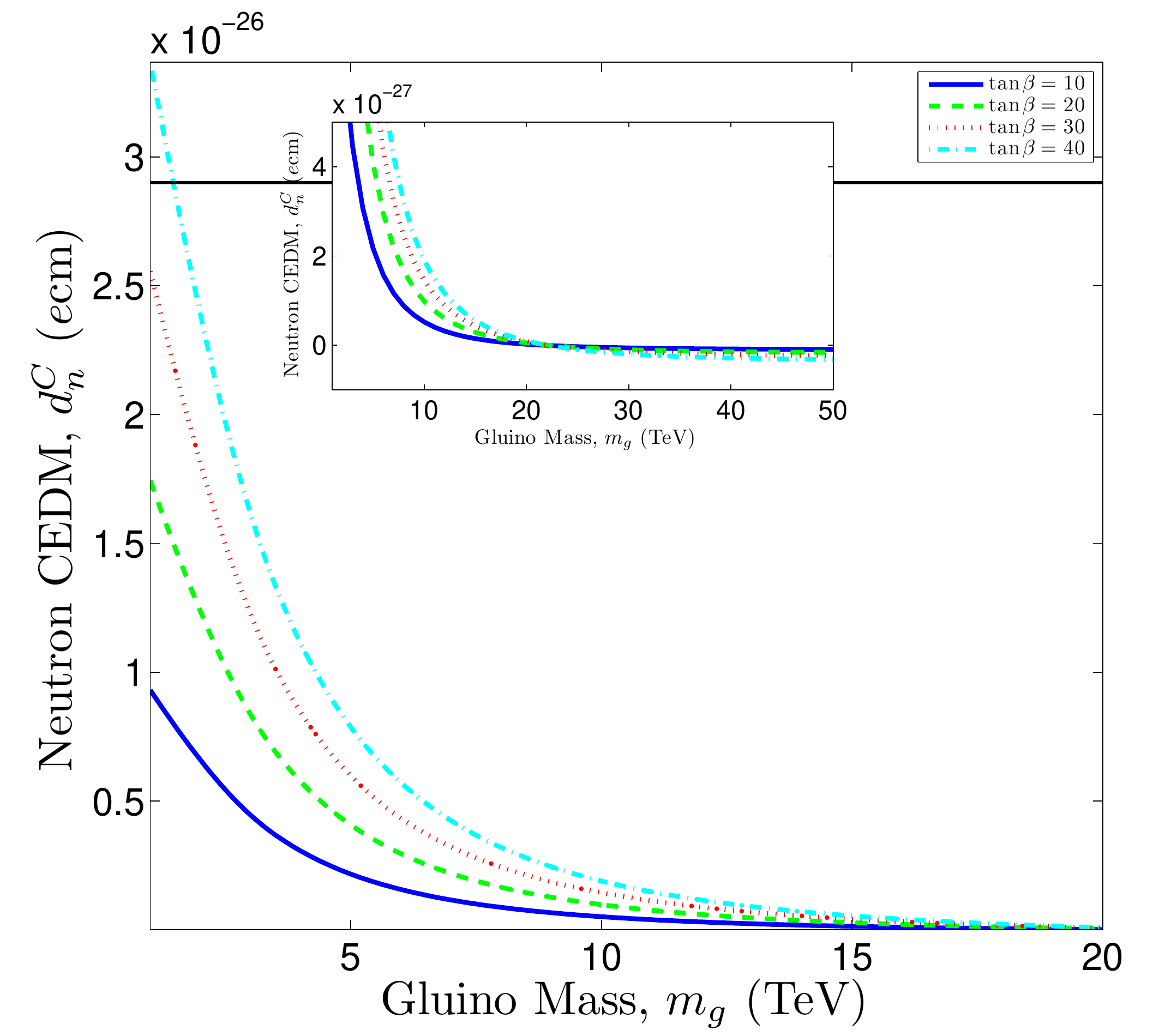}}\hglue5mm}}
\caption{ Variation of neutron CEDM $d^C_{n}$ versus the gluino mass, $m_g$, for four values of $\tan\beta$. From bottom to top at $m_g=5$ TeV, $\tan\beta$ = 10, 20, 30, 40.
Other parameters have the values $|m_{1}|=170$, $|m_{2}|=220$, $|\mu|=450$, $|A^{u}_{0}|=680$, $|A^{d}_{0}|=600$, $m^{u}_{0}=m^{d}_{0}=3700$, $m_{T}=300$, $m_{B}=260$, $m_{4t}=320$, $m_{4b}=280$, $|h_{3}|=1.58$, $|h'_{3}|=6.34\times 10^{-2}$, $|h''_{3}|=1.97\times 10^{-2}$, $|h_{4}|=4.42$, $|h'_{4}|=5.07$, $|h''_{4}|=2.87$, $|h_{5}|=6.6$, $|h'_{5}|=2.67$, $|h''_{5}|=1.86\times 10^{-1}$, $|h_{6}|=|h_{7}|=|h_{8}|=1000$, $\theta_{\mu}=2.6\times10^{-3}$, $\xi_{1}=2\times 10^{-2}$, $\xi_{2}=2\times 10^{-3}$, $\xi_3=1.6$, $\alpha_{A^{u}_{0}}=2\times 10^{-2}$, $\alpha_{A^{d}_{0}}=3.0$, $\chi_{3}=2\times 10^{-2}$, $\chi'_{3}=1\times 10^{-3}$, $\chi''_{3}=4\times 10^{-3}$, $\chi_{4}=7\times 10^{-3}$, $\chi'_{4}=\chi''_{4}=1\times 10^{-3}$, $\chi_{5}=9\times 10^{-3}$, $\chi'_{5}=5\times 10^{-3}$, $\chi''_{5}=2\times 10^{-3}$, $\chi_{6}=\chi_{7}=\chi_{8}=5\times 10^{-3}$. All masses are in GeV and phases in rad.}
\label{fig5a}
\end{center}
\end{figure}

\begin{figure}[H]
\begin{center}
{\rotatebox{0}{\resizebox*{10.5cm}{!}{\includegraphics{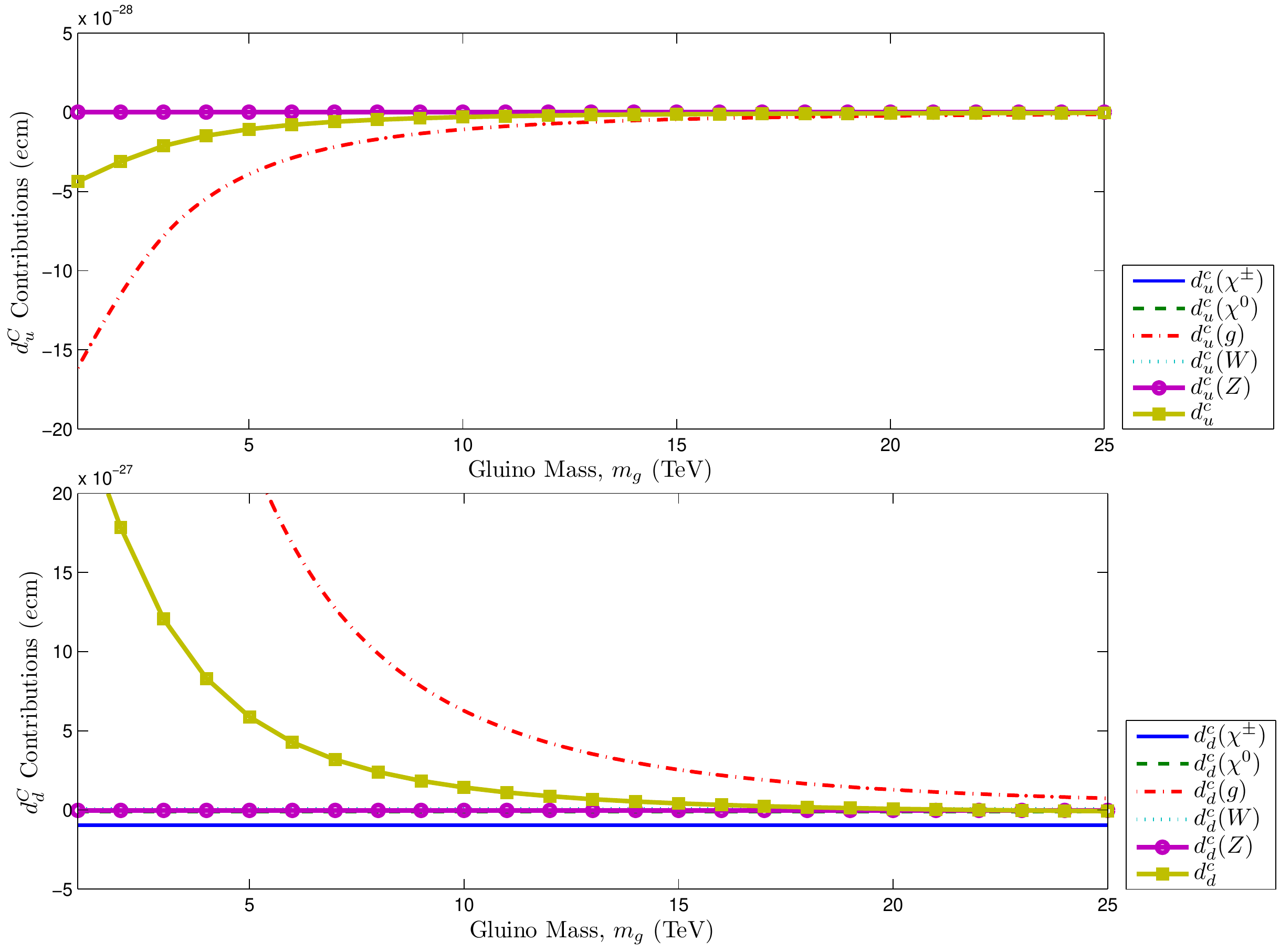}}\hglue5mm}}
\caption{
 Variation of  up and down quark contributions to the neutron CEDM for $\tan\beta=40$. Other parameters are the same as in
 Fig.~\ref{fig5a}. }
\label{fig5b}
\end{center}
\end{figure}

Next we study the dependence of the CEDM on the gluino mass.  This is given in Figs.~\ref{fig5a} and~\ref{fig5b}. Thus in
 Fig.~\ref{fig5a}, the variation of the neutron CEDM, $d^C_n$, is plotted against the gluino mass, $m_g$. It is shown that CEDM values lower than the current experimental upper limit can be obtained in the given parameter space. The neutron CEDM decreases for increasing values of $m_g$, but eventually levels off at around zero for some values of $\tan\beta$. However, for other values of $\tan\beta$, (e.g. $\tan\beta=40$), the neutron CEDM levels off but turns negative.
{This phenomenon can be understood by  analyzing different contributions to the CEDM as  shown in  Fig.~\ref{fig5b}. Specifically one finds that the
negative contribution to the CEDM arises from  the chargino exchange loop contribution, $d^c_d(\chi^\pm)$.
Since we are not applying any GUT constraints, the masses of the chargino and the gluino can be treated
as independent parameters and thus as we increase the gluino mass, the chargino contribution remains unchanged
and eventually dominates as the gluino mass gets large and makes the CEDM negative for $m_g>20$ TeV.
We note here in passing that the W and Z contributions in this region of the parameter space are negligible compared to the other exchange contributions.
}

\begin{figure}[H]
\begin{center}
{\rotatebox{0}{\resizebox*{10cm}{!}{\includegraphics{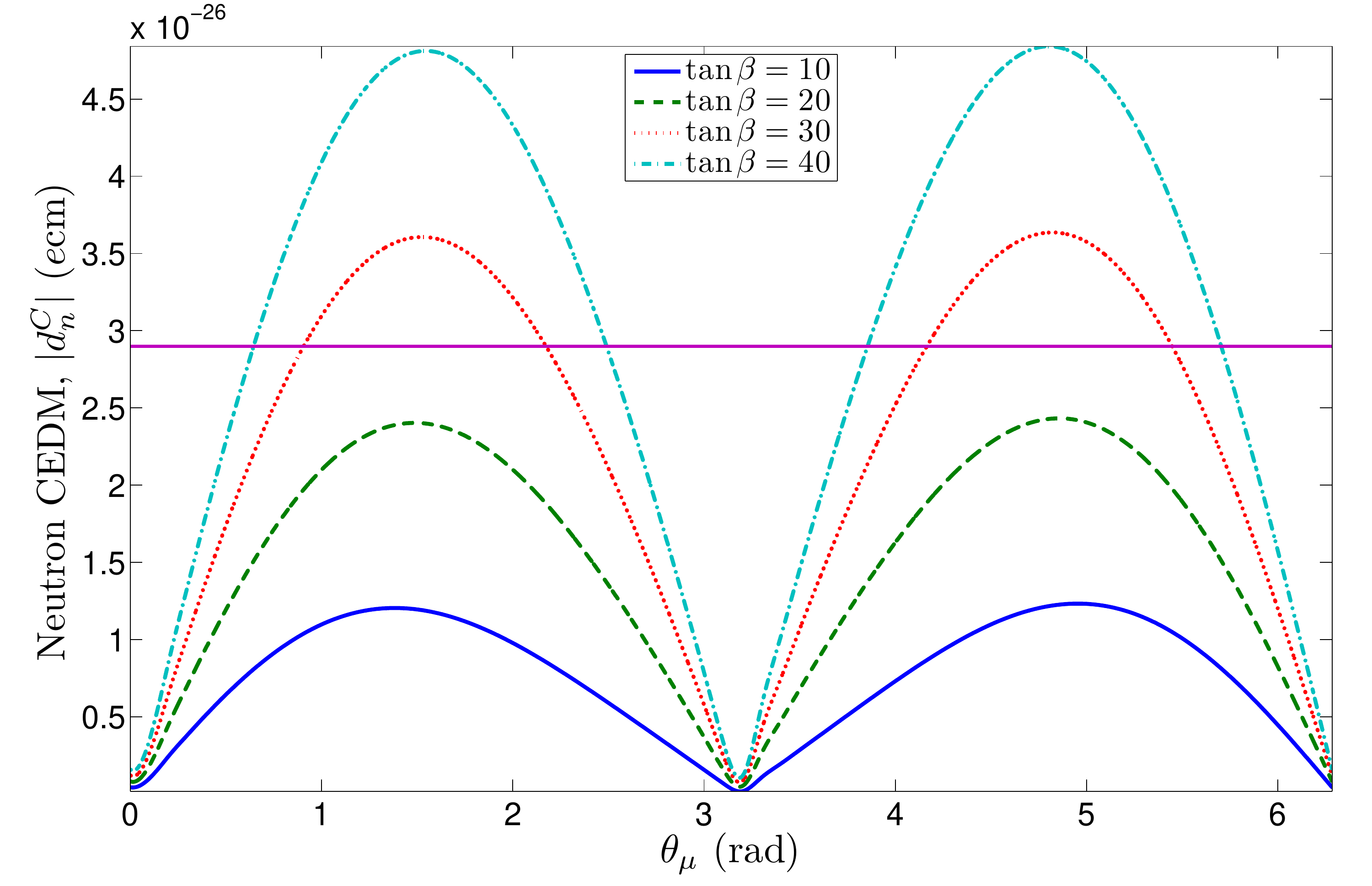}}\hglue5mm}}
\caption{Variation of neutron CEDM $|d^C_{n}|$ versus $\theta_{\mu}$ for four values of $\tan\beta$. From bottom to top at $\theta_{\mu}=1$ rad, $\tan\beta$ = 10, 20, 30, 40. Other parameters have the values $|m_{1}|=170$, $|m_{2}|=220$, $|\mu|=400$, $|A^{u}_{0}|=680$, $|A^{d}_{0}|=600$, $m^{u}_{0}=m^{d}_{0}=8000$, $m_{g}=1000$, $m_{T}=300$, $m_{B}=260$, $m_{4t}=320$, $m_{4b}=280$, $|h_{3}|=1.58$, $|h'_{3}|=6.34\times 10^{-2}$, $|h''_{3}|=1.97\times 10^{-2}$, $|h_{4}|=4.42$, $|h'_{4}|=5.07$, $|h''_{4}|=2.87$, $|h_{5}|=6.6$, $|h'_{5}|=2.67$, $|h''_{5}|=1.86\times 10^{-1}$, $|h_{6}|=|h_{7}|=|h_{8}|=1000$, $\xi_{1}=2\times 10^{-2}$, $\xi_{2}=2\times 10^{-3}$, $\xi_3=2.6$, $\alpha_{A^{u}_{0}}=2\times 10^{-2}$, $\alpha_{A^{d}_{0}}=3.0$, $\chi_{3}=2\times 10^{-2}$, $\chi'_{3}=1\times 10^{-3}$, $\chi''_{3}=4\times 10^{-3}$, $\chi_{4}=7\times 10^{-3}$, $\chi'_{4}=\chi''_{4}=1\times 10^{-3}$, $\chi_{5}=9\times 10^{-3}$, $\chi'_{5}=5\times 10^{-3}$, $\chi''_{5}=2\times 10^{-3}$, $\chi_{6}=\chi_{7}=\chi_{8}=5\times 10^{-3}$. All masses are in GeV and phases in rad.}
\label{fig77}
\end{center}
\end{figure}

{As discussed already, it is of interest to study the dependence of CEDM on the CP phases in the MSSM sector as well as in the new sector.}
Fig.~\ref{fig77} shows the variation of the neutron CEDM versus $\theta_{\mu}$, the phase of $\mu$. The CP phases are the source of the CEDM and the sensitivity that the CEDM shows in response to the variation of $\theta_{\mu}$ is
obvious. The  parameter $\mu$ appears in the chargino and the neutralino mass matrices. It exists also in the squark mass squared  matrices, so one can see that the chargino, the neutralino and the gluino contributions are affected by this parameter and its phase.
The electroweak contributions, i.e, W and Z components are independent of the magnitude and the phase of $\mu$.
 Values of $|d^C_n|$ below the current upper limit can be obtained for several values of $\tan\beta$, whereas values above the limit appear for larger $\tan\beta$. \\

\begin{figure}[H]
\begin{center}
{\rotatebox{0}{\resizebox*{10cm}{!}{\includegraphics{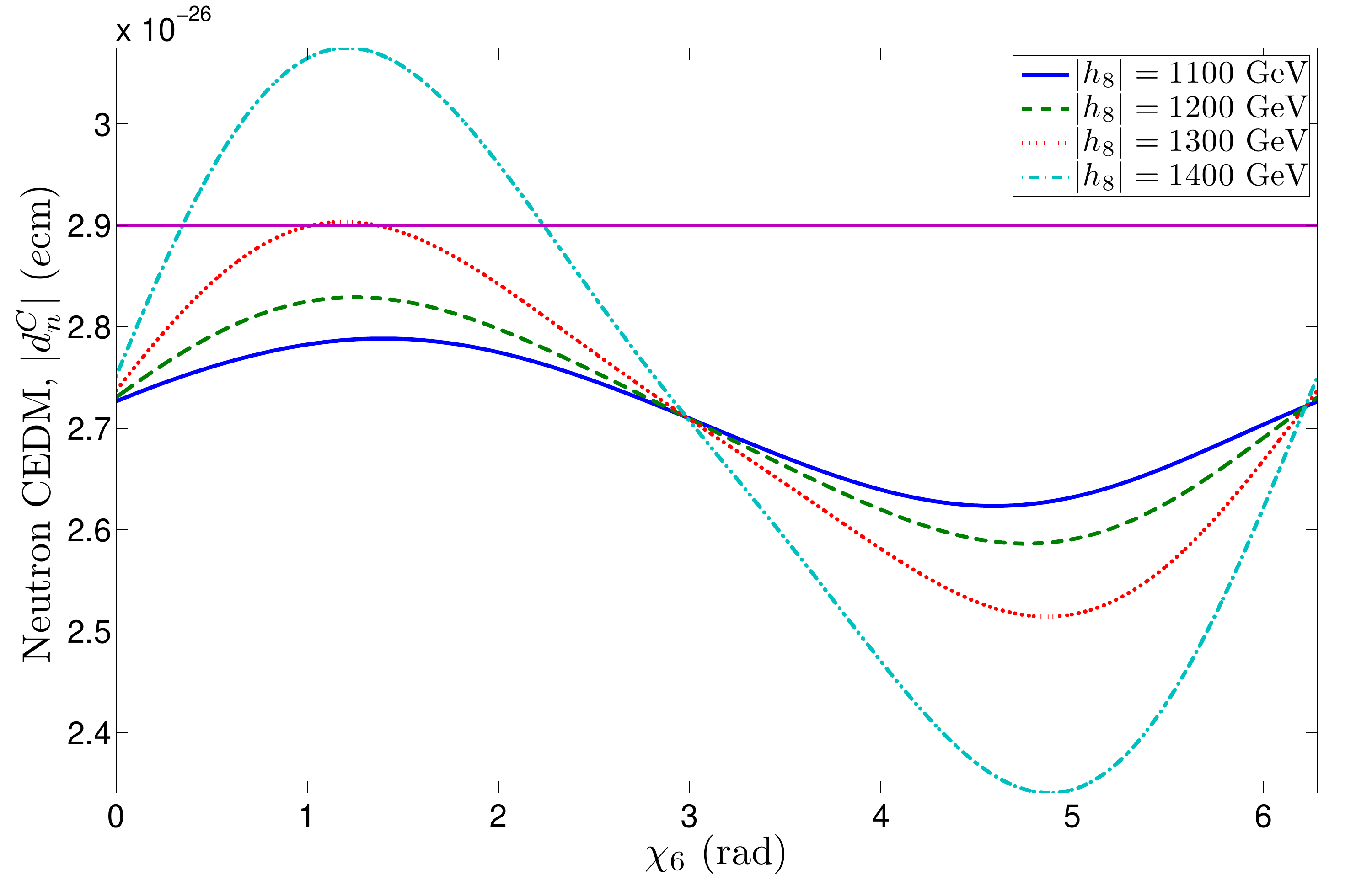}}\hglue5mm}}
\caption{Variation of neutron CEDM $|d^C_{n}|$ versus $\chi_6$, for four values of $|h_8|$. From bottom to top at $\chi_6=1$ rad, $|h_8|$ = 1100, 1200, 1300, 1400 GeV. Other parameters have the values $\tan\beta=34$, $|m_{1}|=185$, $|m_{2}|=220$, $|\mu|=350$, $|A^{u}_{0}|=680$, $|A^{d}_{0}|=600$, $m^{u}_{0}=m^{d}_{0}=3600$, $m_{T}=300$, $m_{B}=260$, $m_g=4000$, $m_{4t}=320$, $m_{4b}=280$, $|h_{3}|=1.58$, $|h'_{3}|=6.34\times 10^{-2}$, $|h''_{3}|=1.97\times 10^{-2}$, $|h_{4}|=4.42$, $|h'_{4}|=5.07$, $|h''_{4}|=2.87$, $|h_{5}|=6.6$, $|h'_{5}|=2.67$, $|h''_{5}|=1.86\times 10^{-1}$, $|h_{6}|=|h_{7}|=1100$, $\theta_{\mu}=0.1$, $\xi_{1}=2\times 10^{-2}$, $\xi_{2}=2\times 10^{-3}$, $\xi_3=3.6$, $\alpha_{A^{u}_{0}}=2\times 10^{-2}$, $\alpha_{A^{d}_{0}}=3.0$, $\chi_{3}=2\times 10^{-2}$, $\chi'_{3}=1\times 10^{-3}$, $\chi''_{3}=4\times 10^{-3}$, $\chi_{4}=7\times 10^{-3}$, $\chi'_{4}=\chi''_{4}=1\times 10^{-3}$, $\chi_{5}=9\times 10^{-3}$, $\chi'_{5}=5\times 10^{-3}$, $\chi''_{5}=2\times 10^{-3}$, $\chi_{7}=\chi_{8}=5\times 10^{-3}$. All masses are in GeV and phases in rad.}
\label{fig88}
\end{center}
\end{figure}

{Next we  investigate the dependence of CEDM on $\chi_6$ which explores a new sector of the theory
as it is the CP phase that arises in interactions involving the mirror quarks and the fourth generation quarks.
An analysis of the dependence of CEDM on $\chi_6$ is exhibited in  Fig.~\ref{fig88}. Aside from $h_6$, other
mass parameters that arise because of the new sector are $h_7$ and $h_8$.  The dependence of the CEDM on
$|h_8|$ is also exhibited in  Fig.~\ref{fig88}. Quite remarkably CEDM is sensitive to both the mass scale and
the phase that enters in the new sector. }

\begin{figure}[H]
\begin{center}
{\rotatebox{0}{\resizebox*{10cm}{!}{\includegraphics{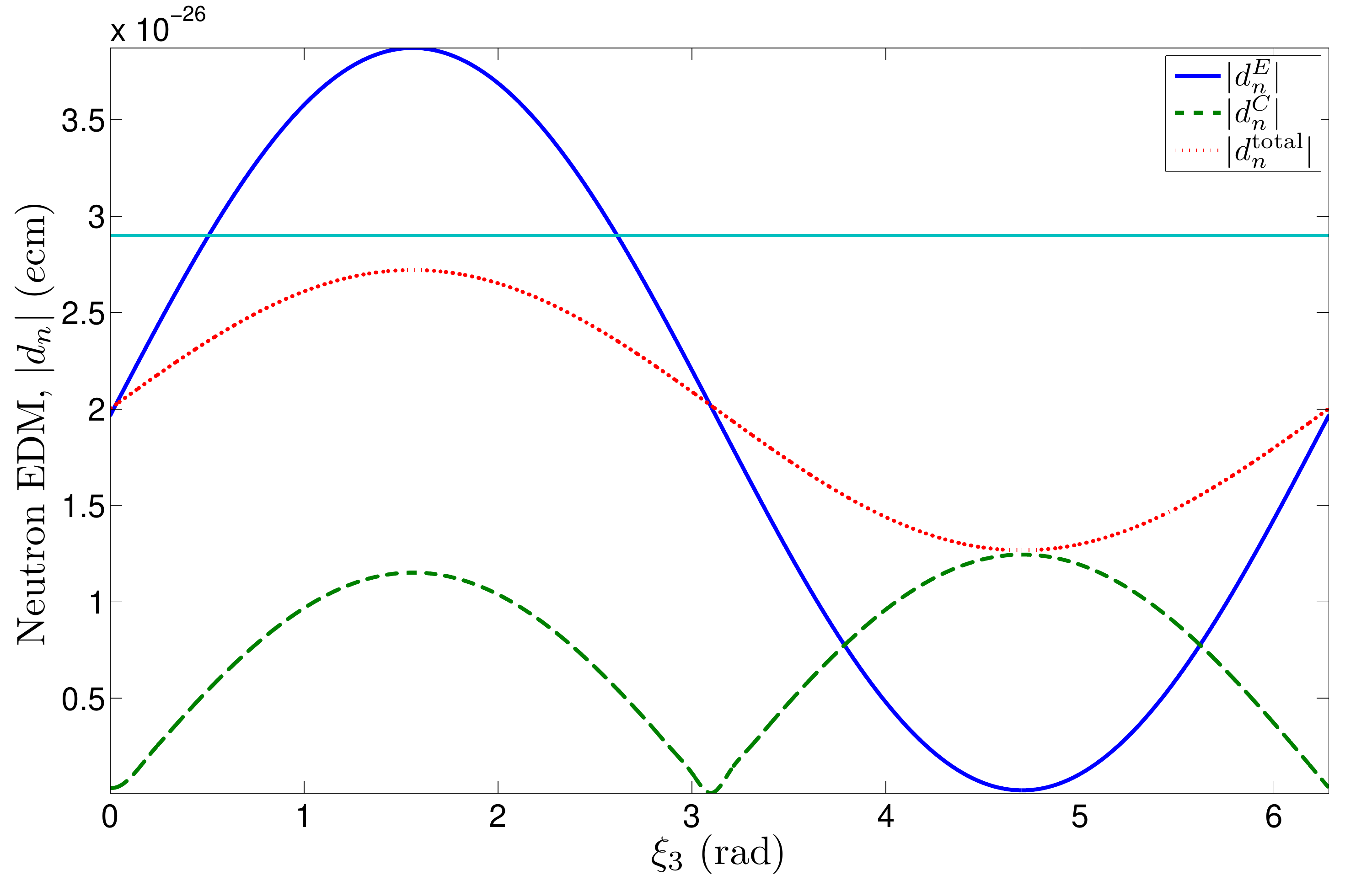}}\hglue5mm}}
\caption{Variation of the neutron EDM, $|d^E_{n}|$ (solid curve), the neutron CEDM, $|d^C_{n}|$ (dashed curve), and the total neutron EDM, $|d^{\rm total}_{n}|$ (dotted curve), versus $\xi_3$, the phase of the gluino mass, for $\tan\beta=40$. Other parameters have the values $|m_{1}|=185$, $|m_{2}|=220$, $|\mu|=400$, $|A^{u}_{0}|=680$, $|A^{d}_{0}|=600$, $m^{u}_{0}=m^{d}_{0}=5000$, $m_{T}=300$, $m_{B}=260$, $m_g=1500$, $m_{4t}=320$, $m_{4b}=280$, $|h_{3}|=1.58$, $|h'_{3}|=6.34\times 10^{-2}$, $|h''_{3}|=1.97\times 10^{-2}$, $|h_{4}|=4.42$, $|h'_{4}|=5.07$, $|h''_{4}|=2.87$, $|h_{5}|=6.6$, $|h'_{5}|=2.67$, $|h''_{5}|=1.86\times 10^{-1}$, $|h_{6}|=|h_{7}|=|h_{8}|=1000$, $\theta_{\mu}=4.7\times 10^{-3}$, $\xi_{1}=2\times 10^{-2}$, $\xi_{2}=2\times 10^{-3}$, $\alpha_{A^{u}_{0}}=2\times 10^{-2}$, $\alpha_{A^{d}_{0}}=3.0$, $\chi_{3}=2\times 10^{-2}$, $\chi'_{3}=1\times 10^{-3}$, $\chi''_{3}=4\times 10^{-3}$, $\chi_{4}=7\times 10^{-3}$, $\chi'_{4}=\chi''_{4}=1\times 10^{-3}$, $\chi_{5}=9\times 10^{-3}$, $\chi'_{5}=5\times 10^{-3}$, $\chi''_{5}=2\times 10^{-3}$, $\chi_{6}=\chi_{7}=\chi_{8}=5\times 10^{-3}$. All masses are in GeV and phases in rad.}\label{fig9}
\end{center}
\end{figure}

Finally, it is of interest to look at the total electric dipole moment obtained by adding the electric and the chromoelectric dipole moments. Fig.~\ref{fig9} shows the variation of the EDM, the CEDM and the total EDM against the gluino phase, $\xi_3$. The analysis of Fig.~\ref{fig9} shows
that while
the EDM may dominate the CEDM for some values of $\xi_3$ the opposite may happen for a different range of $\xi_3$.
The analysis also suggests constructive interference  between EDM and CEDM in some parts of the parameter
space while there is destructive interferences in other parts (i.e., for $0<\xi_3<\pi$)  leading to the cancellations mechanism ~\cite{cancellation1,cancellation2}.

\section{Conclusion \label{sec6}}
In this work we have given an analysis of the chromoelectric dipole moment of quarks and of the neutron
arising in an extension of MSSM where there is an additional vectorlike generation of quarks in the matter sector.
Such an extension brings in new sources of CP violation which can contribute to the chromoelectric dipole
moment  of quarks. {The work presented here consists of analytical results} on five different types of contributions to the chromoelectric dipole moments of quarks which include both non-supersymmetric as well as supersymmetric loop contributions.
In the non-supersymmetric sector we have contributions arising from
the exchanges of the $W$ and $Z$ bosons in the loops, while in the supersymmetric sector we have exchanges
 involving charginos, neutralinos and the gluino in the loop. We have also carried out a detailed numerical
analysis of their relative contributions. {Specifically it is found that there exists strong interference effects
 between the MSSM sector and the vectorlike quark sector which can drastically change both the sign
 and the magnitude of quark EDMs.} We have also investigated the possibility that the neutron EDM can be used
as probe of the TeV scale physics. {These results are of import as future experiment can improve the current
limits up to two orders of magnitude  and thus the quark EDMs provide an important window to new physics
beyond the standard model.}

\noindent
{\bf Acknowledgments:}
PN's research  is  supported in part by the NSF grant PHY-1314774.\\

\noindent
\section{Appendix A: Squark mass  matrices\label{A}}
In this Appendix we give further details of the model discussed in section 2. As discussed in
section 2 we allow for mixing between the vector generation and specifically the mirrors  and the standard three generations of quarks. We also allow for mixing between the mirror generation and the fourth sequential generation assuming $R$ parity conservation
(for a recent review of R parity see \cite{Mohapatra:2015fua}). The superpotential allowing such mixings is given by

\begin{align}
W&=\epsilon_{ij}  [y_{1}  \hat H_1^{i} \hat q_{1L} ^{j}\hat b^c_{1L}
 +y_{1}'  \hat H_2^{j}  \hat q_{1L} ^{i}\hat t^c_{1L}
+y_{2}  \hat H_1^{i} \hat Q^c{^{j}}\hat T_{L}
+y_{2}'  \hat H_2^{j} \hat Q^c{^{i}}\hat B_{L}\nonumber \\
 &+y_{3}  \hat H_1^{i} \hat q_{2L} ^{j}\hat b^c_{2L}
 +y_{3}'  \hat H_2^{j}  \hat q_{2L} ^{i}\hat t^c_{2L}
 +y_{4}  \hat H_1^{i} \hat q_{3L} ^{j}\hat b^c_{3L}
 +y_{4}'  \hat H_2^{j}  \hat q_{3L} ^{i}\hat t^c_{3L}
+y_{5}  \hat H_1^{i} \hat q_{4L} ^{j}\hat b^c_{4L}
+y_{5}'  \hat H_2^{j}  \hat q_{4L} ^{i}\hat t^c_{4L}
] \nonumber \\
&+ h_{3} \epsilon_{ij}  \hat Q^c{^{i}}\hat q_{1L}^{j}+
h_{3}' \epsilon_{ij}  \hat Q^c{^{i}}\hat q_{2L}^{j}+
h_{3}'' \epsilon_{ij}  \hat Q^c{^{i}}\hat q_{3L}^{j}
+h_4 \hat b_{1L}^c \hat B_{L} +h_5 \hat t_{1L}^c \hat T_{L}+h_4'  \hat b_{2L}^c \hat B_{L} \nonumber\\
&+h_5' \hat t_{2L}^c \hat T_{L}
+h_4''  \hat b_{3L}^c \hat B_{L} +h_5'' \hat t_{3L}^c \hat T_{L}
+h_6\epsilon_{ij}Q^i q^j_{4L}+h_7  \hat b_{4L}^c \hat B_{L} +h_8 \hat t_{4L}^c \hat T_{L}
  -\mu \epsilon_{ij} \hat H_1^i \hat H_2^j \ ,
 \label{7w}
\end{align}
Here the couplings are in general complex. Thus, for example,
 $\mu$ is the complex Higgs mixing parameter so that $\mu= |\mu| e^{i\theta_\mu}$.
The mass terms for the ups, mirror ups,  downs and  mirror downs arise from the term
\beq
{\cal{L}}=-\frac{1}{2}\frac{\partial ^2 W}{\partial{A_i}\partial{A_j}}\psi_ i \psi_ j+\text{h.c.},
\label{6}
\eeq
where $\psi$ and $A$ stand for generic two-component fermion and scalar fields.
After spontaneous breaking of the electroweak symmetry, ($\langle H_1^1 \rangle=v_1/\sqrt{2} $ and $\langle H_2^2\rangle=v_2/\sqrt{2}$),
we have the following set of mass terms written in the four-component spinor notation
so that
\beq
-{\cal L}_m= \bar\xi_R^T (M_u) \xi_L +\bar\eta_R^T(M_{d}) \eta_L +\text{h.c.},
\eeq
where the basis vectors are defined in Eq.~(3) and Eq.~(8).

  Next we  consider  the mixing of the down squarks and the charged mirror sdowns.
The mass squared  matrix of the sdown - mirror sdown comes from three sources:  the F term, the
D term of the potential and the soft {SUSY} breaking terms.
Using the  superpotential of  the mass terms arising from it
after the breaking of  the electroweak symmetry are given by
the Lagrangian
\beq
{\cal L}= {\cal L}_F +{\cal L}_D + {\cal L}_{\rm soft}\ ,
\eeq
where   $ {\cal L}_F$ is deduced from $F_i =\partial W/\partial A_i$, and $- {\cal L}_F=V_F=F_i F^{*}_i$ while the ${\cal L}_D$ is given by
\begin{align}
-{\cal L}_D&=\frac{1}{2} m^2_Z \cos^2\theta_W \cos 2\beta \{\tilde t_{ L} \tilde t^*_{ L} -\tilde b_L \tilde b^*_L
+\tilde c_{ L} \tilde c^*_{ L} -\tilde s_L \tilde s^*_L
+\tilde u_{ L} \tilde u^*_{ L} -\tilde d_L \tilde d^*_L
+\tilde t_{4 L} \tilde t^*_{4 L} -\tilde b_{4L} \tilde b^*_{4L}
 \nonumber \\
&+\tilde B_R \tilde B^*_R -\tilde T_R \tilde T^*_R\}
+\frac{1}{2} m^2_Z \sin^2\theta_W \cos 2\beta \{-\frac{1}{3}\tilde t_{ L} \tilde t^*_{ L}
 +\frac{4}{3}\tilde t_{ R} \tilde t^*_{ R}
-\frac{1}{3}\tilde c_{ L} \tilde c^*_{ L}
 +\frac{4}{3}\tilde c_{ R} \tilde c^*_{ R} \nonumber \\
&-\frac{1}{3}\tilde u_{ L} \tilde u^*_{ L}
 +\frac{4}{3}\tilde u_{ R} \tilde u^*_{ R}
+\frac{1}{3}\tilde T_{ R} \tilde T^*_{ R}
 -\frac{4}{3}\tilde T_{ L} \tilde T^*_{ L}
-\frac{1}{3}\tilde b_{ L} \tilde b^*_{ L}
 -\frac{2}{3}\tilde b_{ R} \tilde b^*_{ R}\nonumber\\
&-\frac{1}{3}\tilde s_{ L} \tilde s^*_{ L}
 -\frac{2}{3}\tilde s_{ R} \tilde s^*_{ R}
-\frac{1}{3}\tilde d_{ L} \tilde d^*_{ L}
 -\frac{2}{3}\tilde d_{ R} \tilde d^*_{ R}
+\frac{1}{3}\tilde B_{ R} \tilde B^*_{ R}\nonumber\\
&+\frac{2}{3}\tilde B_{ L} \tilde B^*_{ L}
-\frac{1}{3}\tilde t_{4 L} \tilde t^*_{4 L}
 +\frac{4}{3}\tilde t_{ 4R} \tilde t^*_{4 R}
-\frac{1}{3}\tilde b_{4 L} \tilde b^*_{4 L}
 -\frac{2}{3}\tilde b_{ 4R} \tilde b^*_{ 4R}
\}.
\label{12}
\end{align}
For ${\cal L}_{\rm soft}$ we assume the following form
\begin{align}
-{\cal L}_{\text{soft}}&= M^2_{\tilde 1 L} \tilde q^{k*}_{1 L} \tilde q^k_{1 L}
+ M^2_{\tilde 4 L} \tilde q^{k*}_{4 L} \tilde q^k_{4 L}
+ M^2_{\tilde 2 L} \tilde q^{k*}_{2 L} \tilde q^k_{2 L}
+ M^2_{\tilde 3 L} \tilde q^{k*}_{3 L} \tilde q^k_{3 L}
+ M^2_{\tilde Q} \tilde Q^{ck*} \tilde Q^{ck}
 + M^2_{\tilde t_1} \tilde t^{c*}_{1 L} \tilde t^c_{1 L} \nonumber \\
& + M^2_{\tilde b_1} \tilde b^{c*}_{1 L} \tilde b^c_{1 L}
+ M^2_{\tilde t_2} \tilde t^{c*}_{2 L} \tilde t^c_{2 L}
+ M^2_{\tilde b_4} \tilde b^{c*}_{4 L} \tilde b^c_{4 L}
+ M^2_{\tilde t_4} \tilde t^{c*}_{4 L} \tilde t^c_{4 L}\nonumber\\
&+ M^2_{\tilde t_3} \tilde t^{c*}_{3 L} \tilde t^c_{3 L}
+ M^2_{\tilde b_2} \tilde b^{c*}_{2 L} \tilde b^c_{2 L}
+ M^2_{\tilde b_3} \tilde b^{c*}_{3 L} \tilde b^c_{3 L}
+ M^2_{\tilde B} \tilde B^*_L \tilde B_L
 +  M^2_{\tilde T} \tilde T^*_L \tilde T_L \nonumber \\
&+\epsilon_{ij} \{y_1 A_{b} H^i_1 \tilde q^j_{1 L} \tilde b^c_{1L}
-y_1' A_{t} H^i_2 \tilde q^j_{1 L} \tilde t^c_{1L}
+y_5 A_{4b} H^i_1 \tilde q^j_{4 L} \tilde b^c_{4L}
-y_5' A_{4t} H^i_2 \tilde q^j_{4 L} \tilde t^c_{4L}
+y_3 A_{s} H^i_1 \tilde q^j_{2 L} \tilde b^c_{2L}\nonumber\\
&-y_3' A_{c} H^i_2 \tilde q^j_{2 L} \tilde t^c_{2L}
+y_4 A_{d} H^i_1 \tilde q^j_{3 L} \tilde b^c_{3L}
-y_4' A_{u} H^i_2 \tilde q^j_{3 L} \tilde t^c_{3L}
+y_2 A_{T} H^i_1 \tilde Q^{cj} \tilde T_{L}
-y_2' A_{B} H^i_2 \tilde Q^{cj} \tilde B_{L}
+\text{h.c.}\}\ .
\label{13}
\end{align}
Here $M_{\tilde 1 L},  M_{\tilde T}$, etc are the soft masses and $A_t, A_{b}$, etc are the trilinear couplings.
The trilinear couplings are complex  and we define their phases so that
\begin{gather}
A_b= |A_b| e^{i \alpha_{A_b}} \  ,
 ~~A_{t}=  |A_{t}|
 e^{i\alpha_{A_{t}}} \ ,
  \cdots \ .
\end{gather}
From these terms we construct the scalar mass squared matrices.
Thus we  define the scalar mass squared   matrix $M^2_{\tilde{d}}$  in the basis $(\tilde  b_L, \tilde B_L, \tilde b_R,
\tilde B_R, \tilde s_L, \tilde s_R, \tilde d_L, \tilde d_R,
\tilde b_{4L}, \tilde b_{4R}
)$. We  label the matrix  elements of these as $(M^2_{\tilde d})_{ij}= M^2_{ij}$ which is a hermitian matrix.
We can diagonalize this hermitian mass squared  matrix  by the
 unitary transformation
\begin{gather}
 \tilde D^{d \dagger} M^2_{\tilde d} \tilde D^{d} = diag (M^2_{\tilde d_1},
M^2_{\tilde d_2}, M^2_{\tilde d_3},  M^2_{\tilde d_4},  M^2_{\tilde d_5},  M^2_{\tilde d_6},  M^2_{\tilde d_7},  M^2_{\tilde d_8}
 M^2_{\tilde d_9},  M^2_{\tilde d_{10}}
 )\ .
\end{gather}
Similarly we write the   mass squared   matrix in the up squark sector in the basis $(\tilde  t_{ L}, \tilde T_L,$
$ \tilde t_{ R}, \tilde T_R, \tilde  c_{ L},\tilde c_{ R}, \tilde u_{ L}, \tilde u_{R}
,\tilde t_{4 L}, \tilde t_{4R}
 )$.
 Thus here we denote the up squark  mass squared matrix in the form
$(M^2_{\tilde u})_{ij}=m^2_{ij}$ which is also a hermitian matrix.
We can diagonalize this mass square matrix  by the  unitary transformation
\begin{equation}
 \tilde D^{u\dagger} M^2_{\tilde u} \tilde D^{u} = \text{diag} (M^2_{\tilde u_1}, M^2_{\tilde u_2}, M^2_{\tilde u_3},  M^2_{\tilde u_4},M^2_{\tilde u_5},  M^2_{\tilde u_6}, M^2_{\tilde u_7}, M^2_{\tilde u_8}
, M^2_{\tilde u_9}, M^2_{\tilde u_{10}}
)\ .
\end{equation}

\section{Appendix B:
$W$, $Z$, $\tilde \chi^{\pm}$, $\tilde \chi^0$, $\tilde g$  couplings with quarks \label{B}}

\subsection{W-quark -quark couplings}
The couplings that enter in the W -quark -squark interactions of Eq.~(13)
are defined so that
\begin{align}
G_{L_{ji}}^W&= \frac{g}{\sqrt{2}} [
D^{u*}_{L5j}D^{d}_{L5i}+
D^{u*}_{L4j}D^{d}_{L4i}+
D^{u*}_{L3j}D^{d}_{L3i}+D^{u*}_{L1j}D^{d}_{L1i}],  \\
G_{R_{ji}}^W&= \frac{g}{\sqrt{2}}[D^{u*}_{R2j}D^{d}_{R2i}].
\end{align}

\subsection{Z-quark -quark couplings}
The couplings that enter in the Z -up-quark  interactions of Eq.~(18) are defined so that

\beqn
C_{L_{ji}}^{uZ}=\frac{g}{\cos\theta_{W}} [x_1(
D^{u*}_{L 5j} D^{u}_{L 5i}+
D^{u*}_{L 4j} D^{u}_{L 4i} +D^{u*}_{L 1j} D^{u}_{L 1i}
+D^{u*}_{L 3j} D^{u}_{L 3i})+ y_1D^{u*}_{L 2j} D^{u}_{L 2i}],
\eeqn
and
\beqn
C_{R_{ji}}^{uZ}=\frac{g}{\cos\theta_{W}} [y_1(
D^{u*}_{R 5j} D^{u}_{R 5i} +
D^{u*}_{R 4j} D^{u}_{R 4i} +D^{u*}_{R 1j} D^{u}_{R 1i}
+D^{u*}_{R 3j} D^{u}_{R 3i})+ x_1D^{u*}_{R 2j} D^{u}_{R 2i}],
\eeqn
where
\begin{align}
x_1=\frac{1}{2} -\frac{2}{3} \sin^2 \theta_W,
~~y_1=-\frac{2}{3} \sin^2\theta_W.
\end{align}

The couplings that enter in the Z-down-quark  interactions of Eq.~(20)
 are defined so that
\beqn
C_{L_{ji}}^{dZ}=\frac{g}{\cos\theta_{W}} [x_2(
D^{d*}_{L 5j} D^{d}_{L 5i}+
D^{d*}_{L 4j} D^{d}_{L 4i} +D^{d*}_{L 1j} D^{d}_{L 1i}
+D^{d*}_{L 3j} D^{d}_{L 3i})+ y_2D^{d*}_{L 2j} D^{d}_{L 2i}],
\eeqn
and
\beqn
C_{R_{ji}}^{dZ}=\frac{g}{\cos\theta_{W}} [y_2(
D^{d*}_{R 5j} D^{d}_{R 5i} +
D^{d*}_{R 4j} D^{d}_{R 4i} +D^{d*}_{R 1j} D^{d}_{R 1i}
+D^{d*}_{R 3j} D^{d}_{R 3i})+ x_2D^{d*}_{R 2j} D^{d}_{R 2i}],
\eeqn
where
\begin{align}
x_2 =-\frac{1}{2} +\frac{1}{3} \sin^2 \theta_W, ~~~
y_2 =\frac{1}{3} \sin^2\theta_W.
\end{align}

\subsection{Chargino-quark-squark couplings}
The couplings that enter in the chargino-up-quark-down-squark interactions of Eq.~(23) are given by

\begin{align}
C_{jik}^{Lu}=&g(-\kappa_{u}V^{*}_{i2}D^{u*}_{R4j} \tilde{D}^{d}_{7k}
-\kappa_{c}V^{*}_{i2}D^{u*}_{R3j} \tilde{D}^{d}_{5k}\nonumber\\
&-\kappa_{t}V^{*}_{i2}D^{u*}_{R1j} \tilde{D}^{d}_{1k}
-\kappa_{4t}V^{*}_{i2}D^{u*}_{R5j} \tilde{D}^{d}_{9k}
-\kappa_{B}V^{*}_{i2}D^{u*}_{R2j} \tilde{D}^{d}_{2k}+V^{*}_{i1}D^{u*}_{R2j} \tilde{D}^{d}_{4k}),\\
C_{jik}^{Ru}=&g(-\kappa_{d}U^{}_{i2}D^{u*}_{L4j} \tilde{D}^{d}_{8k}
-\kappa_{s}U^{}_{i2}D^{u*}_{L3j} \tilde{D}^{d}_{6k}
-\kappa_{b}U^{}_{i2}D^{u*}_{L1j} \tilde{D}^{d}_{3k}\nonumber\\
&-\kappa_{4b}U^{}_{i2}D^{u*}_{L5j} \tilde{D}^{d}_{10k}
-\kappa_{T}U^{}_{i2}D^{u*}_{L2j} \tilde{D}^{d}_{4k}\nonumber\\
&+U^{}_{i1}D^{u*}_{L4j} \tilde{D}^{d}_{7k}
+U^{}_{i1}D^{u*}_{L3j} \tilde{D}^{d}_{5k}
+U^{}_{i1}D^{u*}_{L1j} \tilde{D}^{d}_{1k}
+U^{}_{i1}D^{u*}_{L5j} \tilde{D}^{d}_{9k}
),
\end{align}

The couplings that enter in the chargino-down-quark-up-squark interactions of Eq.~(22) are given by

\begin{align}
C_{jik}^{Ld}=&g(-\kappa_{d}U^{*}_{i2}D^{d*}_{R4j} \tilde{D}^{u}_{7k}
-\kappa_{s}U^{*}_{i2}D^{d*}_{R3j} \tilde{D}^{u}_{5k}\nonumber\\
&-\kappa_{b}U^{*}_{i2}D^{d*}_{R1j} \tilde{D}^{u}_{1k}
-\kappa_{4b}U^{*}_{i2}D^{d*}_{R5j} \tilde{D}^{u}_{9k}
-\kappa_{T}U^{*}_{i2}D^{d*}_{R2j} \tilde{D}^{u}_{2k}+U^{*}_{i1}D^{d*}_{R2j} \tilde{D}^{u}_{4k}),\\
C_{jik}^{Rd}=&g(-\kappa_{u}V^{}_{i2}D^{d*}_{L4j} \tilde{D}^{u}_{8k}
-\kappa_{c}V^{}_{i2}D^{d*}_{L3j} \tilde{D}^{u}_{6k}
-\kappa_{t}V^{}_{i2}D^{d*}_{L1j} \tilde{D}^{u}_{3k}
-\kappa_{4t}V^{}_{i2}D^{d*}_{L5j} \tilde{D}^{u}_{10k}\nonumber\\
&-\kappa_{B}V^{}_{i2}D^{d*}_{L2j} \tilde{D}^{u}_{4k}
+V^{}_{i1}D^{d*}_{L4j} \tilde{D}^{u}_{7k}
+V^{}_{i1}D^{d*}_{L3j} \tilde{D}^{u}_{5k}
+V^{}_{i1}D^{d*}_{L1j} \tilde{D}^{u}_{1k}
+V^{}_{i1}D^{d*}_{L5j} \tilde{D}^{u}_{9k}
),
\end{align}
where
\begin{align}
(\kappa_{T},\kappa_{b},\kappa_{s},\kappa_{d},\kappa_{4b})&=\frac{(m_{T},m_{b},m_{s},m_{d}, m_{4b})}{\sqrt{2}m_{W}\cos\beta} , \\~\nonumber\\
(\kappa_{B},\kappa_{t},\kappa_{c},\kappa_{u},\kappa_{4t})&=\frac{(m_{B},m_{t},m_{c},m_{u},m_{4t})}{\sqrt{2}m_{W}\sin\beta} .
\end{align}
and
\begin{equation}
U^* M_C V= {\rm diag} (m_{\tilde \chi_1^-}, m_{\tilde \chi_2^-}).
\label{2.4}
\end{equation}

\subsection{ Neutralino-quark-squark couplings}
   We first give the discuss the couplings that enter the the interactions in the mass diagonal basis involving up quarks,
 up squarks and neutralinos of Eq.~(28).  Here we have,

\begin{align}
C_{uijk}^{'L}=&\sqrt{2}(\alpha_{uj}D^{u *}_{R4i}\tilde{D}^{u}_{7k}-\gamma_{uj}D^{u *}_{R4i}\tilde{D}^{u}_{8k}
+\alpha_{cj}D^{u *}_{R3i}\tilde{D}^{u}_{5k}-\gamma_{cj}D^{u *}_{R3i}\tilde{D}^{u}_{6k}
+\alpha_{tj}D^{u *}_{R1i}\tilde{D}^{u}_{1k}\nonumber\\
&-\gamma_{tj}D^{u *}_{R1i}\tilde{D}^{u}_{3k}
+\alpha_{4tj}D^{u *}_{R5i}\tilde{D}^{u}_{9k}
-\gamma_{4tj}D^{u *}_{R5i}\tilde{D}^{u}_{10k}
+\beta_{Tj}D^{u *}_{R2i}\tilde{D}^{u}_{4k}-\delta_{Tj}D^{u *}_{R2i}\tilde{D}^{u}_{2k}),
\end{align}
\begin{align}
C_{uijk}^{'R}=&\sqrt{2}(\beta_{uj}D^{u *}_{L4i}\tilde{D}^{u}_{7k}-\delta_{uj}D^{u *}_{L4i}\tilde{D}^{u}_{8k}
+\beta_{cj}D^{u *}_{L3i}\tilde{D}^{u}_{5k}-\delta_{cj}D^{u *}_{L3i}\tilde{D}^{u}_{6k}
+\beta_{tj}D^{u *}_{L1i}\tilde{D}^{u}_{1k}\nonumber\\
&-\delta_{tj}D^{u *}_{L1i}\tilde{D}^{u}_{3k}
+\beta_{4tj}D^{u *}_{L5i}\tilde{D}^{u}_{9k}
-\delta_{4tj}D^{u *}_{L5i}\tilde{D}^{u}_{10k}
+\alpha_{Tj}D^{u *}_{L2i}\tilde{D}^{u}_{4k}-\gamma_{Tj}D^{u *}_{L2i}\tilde{D}^{u}_{2k})\,,
\end{align}
where

\begin{align}
\alpha_{T j}&=\frac{gm_{T}X^{*}_{3j}}{2m_{W}\cos\beta} \ ;  && \beta_{Tj}=-\frac{2}{3}eX'_{1j}+\frac{g}{\cos\theta_{W}}X'_{2j}\left(-\frac{1}{2}+\frac{2}{3}\sin^{2}\theta_{W}\right) \\
\gamma_{Tj}&=-\frac{2}{3}eX^{'*}_{1j}+\frac{2}{3}\frac{g\sin^{2}\theta_{W}}{\cos\theta_{W}}X^{'*}_{2j} \  ;  && \delta_{Tj}=-\frac{gm_{T}X_{3j}}{2m_{W}\cos\beta}
\end{align}
and

\begin{align}
\alpha_{4t j}&=\frac{gm_{4t}X_{4j}}{2m_{W}\sin\beta} \ ;  &&
\alpha_{t j}=\frac{gm_{t}X_{4j}}{2m_{W}\sin\beta} \ ;  && \alpha_{cj}=\frac{gm_{c}X_{4j}}{2m_{W}\sin\beta} \ ; && \alpha_{uj}=\frac{gm_{u}X_{4j}}{2m_{W}\sin\beta}  \\
\delta_{4tj}&=-\frac{gm_{4t}X^{*}_{4j}}{2m_{W}\sin\beta} \ ; &&
\delta_{tj}=-\frac{gm_{t}X^{*}_{4j}}{2m_{W}\sin\beta} \ ; && \delta_{cj}=-\frac{gm_{c}X^{*}_{4j}}{2m_{W}\sin\beta} \ ; && \delta_{uj}=-\frac{gm_{u}X^{*}_{4j}}{2m_{W}\sin\beta}
\end{align}
and where

\begin{align}
\beta_{4tj}=
\beta_{tj}=\beta_{cj}=\beta_{uj}&=\frac{2}{3}eX^{'*}_{1j}+\frac{g}{\cos\theta_{W}}X^{'*}_{2j}\left(\frac{1}{2}-\frac{2}{3}\sin^{2}\theta_{W}\right)  \\
\gamma_{4tj}=
\gamma_{tj}=\gamma_{cj}=\gamma_{uj}&=\frac{2}{3}eX'_{1j}-\frac{2}{3}\frac{g\sin^{2}\theta_{W}}{\cos\theta_{W}}X'_{2j}
\end{align}

 Similarly for the couplings that enter the the interactions in the mass diagonal basis involving down quarks,
 down squarks and neutralinos of Eq.~(29) we have

 \begin{align}
C_{dijk}^{'L}=&\sqrt{2}(\alpha_{dj}D^{d *}_{R4i}\tilde{D}^{d}_{7k}-\gamma_{dj}D^{d *}_{R4i}\tilde{D}^{d}_{8k}
+\alpha_{sj}D^{d *}_{R3i}\tilde{D}^{d}_{5k}-\gamma_{sj}D^{d *}_{R3i}\tilde{D}^{d}_{6k}
+\alpha_{bj}D^{d *}_{R1i}\tilde{D}^{d}_{1k}-\gamma_{bj}D^{d *}_{R1i}\tilde{D}^{d}_{3k} \nonumber\\
&+\alpha_{4bj}D^{d *}_{R5i}\tilde{D}^{d}_{9k}-\gamma_{4bj}D^{d *}_{R5i}\tilde{D}^{d}_{10k}
+\beta_{Bj}D^{d *}_{R2i}\tilde{D}^{d}_{4k}-\delta_{Bj}D^{d *}_{R2i}\tilde{D}^{d}_{2k}),
\end{align}

and
\begin{align}
C_{dijk}^{'R}=&\sqrt{2}(\beta_{dj}D^{d *}_{L4i}\tilde{D}^{d}_{7k}-\delta_{dj}D^{d *}_{L4i}\tilde{D}^{d}_{8k}
+\beta_{sj}D^{d *}_{L3i}\tilde{D}^{d}_{5k}-\delta_{sj}D^{d *}_{L3i}\tilde{D}^{d}_{6k}
+\beta_{bj}D^{d *}_{L1i}\tilde{D}^{d}_{1k}-\delta_{bj}D^{d *}_{L1i}\tilde{D}^{d}_{3k} \nonumber\\
&+\beta_{4bj}D^{d *}_{L5i}\tilde{D}^{d}_{9k}-\delta_{4bj}D^{d *}_{L5i}\tilde{D}^{d}_{10k}
+\alpha_{Bj}D^{d *}_{L2i}\tilde{D}^{d}_{4k}-\gamma_{Bj}D^{d *}_{L2i}\tilde{D}^{d}_{2k}),
\end{align}
where

\begin{align}
\alpha_{B j}&=\frac{gm_{B}X^{*}_{4j}}{2m_{W}\sin\beta} \ ;  && \beta_{Bj}=\frac{1}{3}eX'_{1j}+\frac{g}{\cos\theta_{W}}X'_{2j}\left(\frac{1}{2}-\frac{1}{3}\sin^{2}\theta_{W}\right) \\
\gamma_{Bj}&=\frac{1}{3}eX^{'*}_{1j}-\frac{1}{3}\frac{g\sin^{2}\theta_{W}}{\cos\theta_{W}}X^{'*}_{2j} \  ;  && \delta_{Bj}=-\frac{gm_{B}X_{4j}}{2m_{W}\sin\beta}
\end{align}
and

\begin{align}
\alpha_{4b j}&=\frac{gm_{4b}X_{3j}}{2m_{W}\cos\beta} \ ;  &&
\alpha_{b j}=\frac{gm_{b}X_{3j}}{2m_{W}\cos\beta} \ ;  && \alpha_{sj}=\frac{gm_{s}X_{3j}}{2m_{W}\cos\beta} \ ; && \alpha_{dj}=\frac{gm_{d}X_{3j}}{2m_{W}\cos\beta}  \\
\delta_{4bj}&=-\frac{gm_{4b}X^{*}_{3j}}{2m_{W}\cos\beta} \ ; &&
\delta_{bj}=-\frac{gm_{b}X^{*}_{3j}}{2m_{W}\cos\beta} \ ; && \delta_{sj}=-\frac{gm_{s}X^{*}_{3j}}{2m_{W}\cos\beta} \ ; && \delta_{dj}=-\frac{gm_{d}X^{*}_{3j}}{2m_{W}\cos\beta}
\end{align}
and where

\begin{align}
\beta_{4bj}=
\beta_{bj}=\beta_{sj}=\beta_{dj}&=-\frac{1}{3}eX^{'*}_{1j}+\frac{g}{\cos\theta_{W}}X^{'*}_{2j}\left(-\frac{1}{2}+\frac{1}{3}\sin^{2}\theta_{W}\right)  \\
\gamma_{4bj}=
\gamma_{bj}=\gamma_{sj}=\gamma_{dj}&=-\frac{1}{3}eX'_{1j}+\frac{1}{3}\frac{g\sin^{2}\theta_{W}}{\cos\theta_{W}}X'_{2j}
\end{align}
Here $X'$ are defined by

\begin{align}
X'_{1i}&=X_{1i}\cos\theta_{W}+X_{2i}\sin\theta_{W}  \\
X'_{2i}&=-X_ {1i}\sin\theta_{W}+X_{2i}\cos\theta_{W},
\end{align}
where $X$ diagonalizes the neutralino mass matrix and is defined by

 \begin{equation}
X^T M_{\chi^0} X= {\rm diag} \left( m_{\tilde \chi_1^0},  m_{\tilde \chi_2^0}, m_{\tilde \chi_3^0}, m_{\tilde \chi_4^0}\right).
\label{2.8}
\end{equation}

\subsection{Gluino-quark-squark-couplings}
The couplings that enter in the gluino-quark-squark interactions of Eq.~(32) are given by

\beqn
C_{L_{lm}}=(D^{q*}_{R2l} \tilde{D}^{q}_{4m}
-D^{q*}_{R5l} \tilde{D}^{q}_{10m}
-D^{q*}_{R4l} \tilde{D}^{q}_{8m}-
D^{q*}_{R3l} \tilde{D}^{q}_{6m}
-D^{q*}_{R1l} \tilde{D}^{q}_{3m}
)e^{-i\xi_3/2},
\eeqn
and
\beqn
C_{R_{lm}}=(D^{q*}_{L4l} \tilde{D}^{q}_{7m}
+D^{q*}_{L5l} \tilde{D}^{q}_{9m}
+D^{q*}_{L3l} \tilde{D}^{q}_{5m}+
D^{q*}_{L1l} \tilde{D}^{q}_{1m}
-D^{q*}_{L2l} \tilde{D}^{q}_{2m}
)e^{i\xi_3/2},
\eeqn
where $\xi_3$ is the phase of the gluino mass.

\noindent
\section{Appendix C: Mass squared matrices for the scalars  \label{sec9}}

We define the scalar mass squared   matrix $M^2_{\tilde{d}}$  in the basis $(\tilde  b_L, \tilde B_L, \tilde b_R,
\tilde B_R, \tilde s_L, \tilde s_R, \tilde d_L, \tilde d_R,
 \tilde b_{4L}, \tilde b_{4R}
)$. We  label the matrix  elements of these as $(M^2_{\tilde d})_{ij}= M^2_{ij}$ where the elements of the matrix are given by
\begin{align}
M^2_{11}&= M^2_{\tilde 1 L}+\frac{v^2_1|y_1|^2}{2} +|h_3|^2 -m^2_Z \cos 2 \beta \left(\frac{1}{2}-\frac{1}{3}\sin^2\theta_W\right), \nonumber\\
M^2_{22}&=M^2_{\tilde B}+\frac{v^2_2|y'_2|^2}{2}+|h_4|^2 +|h'_4|^2+|h''_4|^2
+|h_7|^2
+\frac{1}{3}m^2_Z \cos 2 \beta \sin^2\theta_W, \nonumber\\
M^2_{33}&= M^2_{\tilde b_1}+\frac{v^2_1|y_1|^2}{2} +|h_4|^2 -\frac{1}{3}m^2_Z \cos 2 \beta \sin^2\theta_W, \nonumber\\
M^2_{44}&=  M^2_{\tilde Q}+\frac{v^2_2|y'_2|^2}{2} +|h_3|^2 +|h'_3|^2+|h''_3|^2
+|h_6|^2
 +m^2_Z \cos 2 \beta \left(\frac{1}{2}-\frac{1}{3}\sin^2\theta_W\right), \nonumber
\end{align}
\begin{align}
M^2_{55}&=M^2_{\tilde 2 L} +\frac{v^2_1|y_3|^2}{2} +|h'_3|^2 -m^2_Z \cos 2 \beta \left(\frac{1}{2}-\frac{1}{3}\sin^2\theta_W\right), \nonumber\\
M^2_{66}&= M^2_{\tilde b_2}+\frac{v^2_1|y_3|^2}{2}+|h'_4|^2  -\frac{1}{3}m^2_Z \cos 2 \beta \sin^2\theta_W,\nonumber\\
M^2_{77}&=M^2_{\tilde 3 L}+\frac{v^2_1|y_4|^2}{2}+|h''_3|^2-m^2_Z \cos 2 \beta \left(\frac{1}{2}-\frac{1}{3}\sin^2\theta_W\right),  \nonumber\\
M^2_{88}&= M^2_{\tilde b_3}+\frac{v^2_1|y_4|^2}{2}+|h''_4|^2   -\frac{1}{3}m^2_Z \cos 2 \beta \sin^2\theta_W\ . \nonumber\\
M^2_{99}&=M^2_{\tilde 4 L}+\frac{v^2_1|y_5|^2}{2}+|h_6|^2-m^2_Z \cos 2 \beta \left(\frac{1}{2}-\frac{1}{3}\sin^2\theta_W\right)\nonumber\\
M^2_{1010}&= M^2_{\tilde b_4}+\frac{v^2_1|y_5|^2}{2}+|h_7|^2   -\frac{1}{3}m^2_Z \cos 2 \beta \sin^2\theta_W\ . \nonumber\\
\end{align}

\begin{align}
M^2_{12}=M^{2*}_{21}=\frac{ v_2 y'_2h^*_3}{\sqrt{2}} +\frac{ v_1 h_4 y^*_1}{\sqrt{2}} ,
M^2_{13}=M^{2*}_{31}=\frac{y^*_1}{\sqrt{2}}(v_1 A^*_{b} -\mu v_2),
M^2_{14}=M^{2*}_{41}=0,\nonumber\\
 M^2_{15} =M^{2*}_{51}=h'_3 h^*_3,
 M^{2}_{16}= M^{2*}_{61}=0,  M^{2}_{17}= M^{2*}_{71}=h''_3 h^*_3,  M^{2*}_{18}= M^{2*}_{81}=0,
M^{2}_{19}=M^{2*}_{91}=h^*_3 h_6,
\nonumber\\
M^{2}_{110}=M^{2*}_{101}=0,
M^2_{23}=M^{2*}_{32}=0,
M^2_{24}=M^{2*}_{42}=\frac{y'^*_2}{\sqrt{2}}(v_2 A^*_{B} -\mu v_1),  M^2_{25} = M^{2*}_{52}= \frac{ v_2 h'_3y'^*_2}{\sqrt{2}} +\frac{ v_1 y_3 h^*_4}{\sqrt{2}} ,\nonumber\\
 M^2_{26}=M^{2*}_{62}=0,  M^2_{27} =M^{2*}_{72}=  \frac{ v_2 h''_3y'^*_2}{\sqrt{2}} +\frac{ v_1 y_4 h''^*_4}{\sqrt{2}},  M^2_{28} =M^{2*}_{82}=0, \nonumber\\
 M^2_{29} =M^{2*}_{92}=  \frac{ v_1 h^*_7y_5}{\sqrt{2}} +\frac{ v_2 y'^*_2 h_6}{\sqrt{2}},
M^{2}_{210}=M^{2*}_{102}=0,\nonumber\\
M^2_{34}=M^{2*}_{43}= \frac{ v_2 h_4 y'^*_2}{\sqrt{2}} +\frac{ v_1 y_1 h^*_3}{\sqrt{2}}, M^2_{35} =M^{2*}_{53} =0, M^2_{36} =M^{2*}_{63}=h_4 h'^*_4,\nonumber\\
 M^2_{37} =M^{2*}_{73} =0,  M^2_{38} =M^{2*}_{83} =h_4 h''^*_4,\nonumber\\
M^{2}_{39}=M^{2*}_{93}=0,
M^{2}_{310}=M^{2*}_{103}=h_4 h^*_7,\nonumber\\
M^2_{45}=M^{2*}_{54}=0, M^2_{46}=M^{2*}_{64}=\frac{ v_2 y'_2 h'^*_4}{\sqrt{2}} +\frac{ v_1 h'_3 y^*_3}{\sqrt{2}}, \nonumber\\
 M^2_{47} =M^{2*}_{74}=0,  M^2_{48} =M^{2*}_{84}=  \frac{ v_2 y'_2h''^*_4}{\sqrt{2}} +\frac{ v_1 h''_3 y^*_4}{\sqrt{2}},\nonumber\\
M^{2}_{49}=M^{2*}_{94}=0,
 M^2_{410} =M^{2*}_{104}=  \frac{ v_2 y'_2h^*_7}{\sqrt{2}} +\frac{ v_1 h_6 y^*_5}{\sqrt{2}},\nonumber\\
M^2_{56}=M^{2*}_{65}=\frac{y^*_3}{\sqrt{2}}(v_1 A^*_{s} -\mu v_2),
 M^2_{57} =M^{2*}_{75}=h''_3 h'^*_3,  \nonumber\\
 M^2_{58} =M^{2*}_{85}=0,
M^2_{59} =M^{2*}_{95}=h'^*_3 h_6,
 M^2_{510} =M^{2*}_{105}=0,
  M^2_{67} =M^{2*}_{76}=0,\nonumber\\
 M^2_{68} =M^{2*}_{86}=h'_4 h''^*_4,
 M^2_{69} =M^{2*}_{96}=0,
 M^2_{610} =M^{2*}_{106}=h'_4 h^*_7,
  M^2_{78}=M^{2*}_{87}=\frac{y^*_4}{\sqrt{2}}(v_1 A^*_{d} -\mu v_2)\ . \nonumber\\
 M^2_{79} =M^{2*}_{97}=h''^*_3 h_6,
 M^2_{710} =M^{2*}_{107}=0\nonumber\\
 M^2_{89} =M^{2*}_{98}=0,
 M^2_{810} =M^{2*}_{108}=h''_4 h^*_7,
  M^2_{910}=M^{2*}_{109}=\frac{y^*_5}{\sqrt{2}}(v_1 A^*_{4b} -\mu v_2)\ . \nonumber
\label{14}
\end{align}

We can diagonalize this hermitian mass squared  matrix  by the
 unitary transformation
\begin{gather}
 \tilde D^{d \dagger} M^2_{\tilde d} \tilde D^{d} = diag (M^2_{\tilde d_1},
M^2_{\tilde d_2}, M^2_{\tilde d_3},  M^2_{\tilde d_4},  M^2_{\tilde d_5},  M^2_{\tilde d_6},  M^2_{\tilde d_7},  M^2_{\tilde d_8},
M^2_{\tilde d_9},
M^2_{\tilde d_{10}}
 )\ .
\end{gather}


 Next we write the   mass$^2$  matrix in the sups sector the basis $(\tilde  t_{ L}, \tilde T_L,$
$ \tilde t_{ R}, \tilde T_R, \tilde  c_{ L},\tilde c_{ R}, \tilde u_{ L}, \tilde u_{R},
 \tilde t _{4 L}, \tilde t_{4R}
 )$.
 Thus here we denote the sups mass$^2$ matrix in the form
$(M^2_{\tilde u})_{ij}=m^2_{ij}$ where

\begin{align}
m^2_{11}&= M^2_{\tilde 1 L}+\frac{v^2_2|y'_1|^2}{2} +|h_3|^2 +m^2_Z \cos 2 \beta \left(\frac{1}{2}-\frac{2}{3}\sin^2\theta_W\right), \nonumber\\
m^2_{22}&=M^2_{\tilde T}+\frac{v^2_1|y_2|^2}{2}+|h_5|^2 +|h'_5|^2+|h''_5|^2
+|h_8|^2
 -\frac{2}{3}m^2_Z \cos 2 \beta \sin^2\theta_W, \nonumber\\
m^2_{33}&= M^2_{\tilde t_1}+\frac{v^2_2|y'_1|^2}{2} +|h_5|^2 +\frac{2}{3}m^2_Z \cos 2 \beta \sin^2\theta_W, \nonumber\\
m^2_{44}&=  M^2_{\tilde Q}+\frac{v^2_1|y_2|^2}{2} +|h_3|^2 +|h'_3|^2+|h''_3|^2
+|h_6|^2
 -m^2_Z \cos 2 \beta \left(\frac{1}{2}-\frac{2}{3}\sin^2\theta_W\right), \nonumber
\end{align}
\begin{align}
m^2_{55}&=M^2_{\tilde 2 L} +\frac{v^2_2|y'_3|^2}{2} +|h'_3|^2 +m^2_Z \cos 2 \beta \left(\frac{1}{2}-\frac{2}{3}\sin^2\theta_W\right), \nonumber\\
m^2_{66}&= M^2_{\tilde t_2}+\frac{v^2_2|y'_3|^2}{2}+|h'_5|^2  +\frac{2}{3}m^2_Z \cos 2 \beta \sin^2\theta_W,\nonumber\\
m^2_{77}&=M^2_{\tilde 3 L}+\frac{v^2_2|y'_4|^2}{2}+|h''_3|^2+m^2_Z \cos 2 \beta \left(\frac{1}{2}-\frac{2}{3}\sin^2\theta_W\right),  \nonumber\\
m^2_{88}&= M^2_{\tilde t_3}+\frac{v^2_2|y'_4|^2}{2}+|h''_5|^2   +\frac{2}{3}m^2_Z \cos 2 \beta \sin^2\theta_W,\nonumber\\
m^2_{99}&=M^2_{\tilde 4 L}+\frac{v^2_2|y'_5|^2}{2}+|h_6|^2+m^2_Z \cos 2 \beta \left(\frac{1}{2}-\frac{2}{3}\sin^2\theta_W\right),  \nonumber\\
m^2_{1010}&= M^2_{\tilde t_4}+\frac{v^2_2|y'_5|^2}{2}+|h_8|^2   +\frac{2}{3}m^2_Z \cos 2 \beta \sin^2\theta_W.\
\nonumber
\end{align}

\begin{align}
m^2_{12}&=m^{2*}_{21}=-\frac{ v_1 y_2h^*_3}{\sqrt{2}} +\frac{ v_2 h_5 y'^*_1}{\sqrt{2}} ,
m^2_{13}=m^{2*}_{31}=\frac{y'^*_1}{\sqrt{2}}(v_2 A^*_{t} -\mu v_1),
m^2_{14}=m^{2*}_{41}=0,\nonumber\\
 m^2_{15} &=m^{2*}_{51}=h'_3 h^*_3,
 m^{2*}_{16}= m^{2*}_{61}=0,  m^{2*}_{17}= m^{2*}_{71}=h''_3 h^*_3,  m^{2*}_{18}= m^{2*}_{81}=0,\nonumber\\
m^2_{23}&=m^{2*}_{32}=0,
m^2_{24}=m^{2*}_{42}=\frac{y^*_2}{\sqrt{2}}(v_1 A^*_{T} -\mu v_2),  m^2_{25} = m^{2*}_{52}= -\frac{ v_1 h'_3y^*_2}{\sqrt{2}} +\frac{ v_2 y'_3 h'^*_5}{\sqrt{2}} ,\nonumber\\
 m^2_{26} &=m^{2*}_{62}=0,  m^2_{27} =m^{2*}_{72}=  -\frac{ v_1 h''_3y^*_2}{\sqrt{2}} +\frac{ v_2 y'_4 h''^*_5}{\sqrt{2}},  m^2_{28} =m^{2*}_{82}=0, \nonumber\\
m^2_{34}&=m^{2*}_{43}= \frac{ v_1 h_5 y^*_2}{\sqrt{2}} -\frac{ v_2 y'_1 h^*_3}{\sqrt{2}}, m^2_{35} =m^{2*}_{53} =0, m^2_{36} =m^{2*}_{63}=h_5 h'^*_5,\nonumber\\
 m^2_{37} &=m^{2*}_{73} =0,  m^2_{38} =m^{2*}_{83} =h_5 h''^*_5,\nonumber\\
m^2_{45}&=m^{2*}_{54}=0, m^2_{46}=m^{2*}_{64}=-\frac{y'^*_3 v_2 h'_3}{\sqrt{2}}+\frac{v_1 y_2 h'^*_5}{\sqrt{2}},
\nonumber\\
m^2_{47}&=m^{2*}_{74}=0,
m^2_{48}=m^{2*}_{84}=\frac{v_1 y_2 h''^*_5}{\sqrt{2}}-\frac{v_2 y'^*_4 h''_3}{\sqrt{2}},\nonumber\\
 m^2_{56}&=m^{2*}_{65}=\frac{y'^*_3}{\sqrt{2}}(v_2 A^*_{c}-\mu v_1), \nonumber\\
m^2_{57}&=m^{2*}_{75}= h''_3 h'^*_3, m^2_{58}=m^{2*}_{85}=0, \nonumber\\
m^2_{67}&=m^{2*}_{76}=0, m^2_{68}=m^{2*}_{86}= h'_5 h''^*_5, \nonumber\\
m^2_{78}&=m^{2*}_{87}=\frac{y'^*_4}{\sqrt{2}}(v_2 A^*_{u}-\mu v_1),\nonumber\\
m^2_{19}&=m^{2*}_{91}=h_6 h^*_3, m^2_{110}=m^{2*}_{101}=0, \nonumber\\
m^2_{29}&=m^{2*}_{92}=-\frac{y^*_2 v_1 h_6}{\sqrt{2}}+\frac{v_2 y^*_5 h_8}{\sqrt{2}},\nonumber\\
m^2_{210}&=m^{2*}_{102}=0, m^2_{39}=m^{2*}_{93}=0,\nonumber\\
m^2_{310}&=m^{2*}_{103}=h_5 h^*_8,\nonumber\\
m^2_{49}&=m^{2*}_{94}=0, m^2_{410}=m^{2*}_{104}=
-\frac{y'^*_5 v_2 h_6}{\sqrt{2}}+\frac{v_1 y_2 h^*_8}{\sqrt{2}},\nonumber\\
m^2_{59}&=m^{2*}_{95}=h_6 h'^*_3, m^2_{510}=m^{2*}_{105}=0\nonumber\\
m^2_{69}&=m^{2*}_{96}=0, m^2_{610}=m^{2*}_{106}= h'_5 h^*_8 \nonumber\\
m^2_{79}&=m^{2*}_{97}=h_6 h''^*_3, m^2_{710}=m^{2*}_{107}=0, \nonumber\\
m^2_{89}&=m^{2*}_{98}=0, m^2_{810}=m^{2*}_{108}=h''_5 h^*_8, \nonumber\\
 m^2_{910}&=m^{2*}_{109}=\frac{y'^*_5}{\sqrt{2}}(v_2 A^*_{4t}-\mu v_1)
\end{align}

We can diagonalize the sneutrino mass square matrix  by the  unitary transformation
\begin{equation}
 \tilde D^{u\dagger} M^2_{\tilde u} \tilde D^{u} = \text{diag} (M^2_{\tilde u_1}, M^2_{\tilde u_2}, M^2_{\tilde u_3},  M^2_{\tilde u_4},M^2_{\tilde u_5},  M^2_{\tilde u_6}, M^2_{\tilde u_7}, M^2_{\tilde u_8}
 M^2_{\tilde u_9}, M^2_{\tilde u_{10}}
)\ .
\end{equation}

\end{document}